\documentclass[10pt,aps,pre,twocolumn,superscriptaddress]{revtex4-1}

\usepackage{amsmath}
\usepackage{textcomp}
\usepackage{amssymb} 
\usepackage{graphicx}
\usepackage{tabularx}
\usepackage{placeins}

\usepackage[usenames]{color}

\makeatletter
\newcommand*{\rom}[1]{\expandafter\@slowromancap\romannumeral #1@} 
\makeatother

\begin{document}

\title{Finite size effects in critical fiber networks}

\author{Sadjad Arzash}
\affiliation{Department of Chemical \& Biomolecular Engineering, Rice University, Houston, TX 77005}
\affiliation{Center for Theoretical Biological Physics, Rice University, Houston, TX 77030}
\author{Jordan L. Shivers}
\affiliation{Department of Chemical \& Biomolecular Engineering, Rice University, Houston, TX 77005}
\affiliation{Center for Theoretical Biological Physics, Rice University, Houston, TX 77030}
\author{Fred C.\ MacKintosh}
\affiliation{Department of Chemical \& Biomolecular Engineering, Rice University, Houston, TX 77005}
\affiliation{Center for Theoretical Biological Physics, Rice University, Houston, TX 77030}
\affiliation{Departments of Chemistry and Physics \& Astronomy, Rice University, Houston, TX 77005}

\begin{abstract}
Fibrous networks such as collagen are common in physiological systems. One important function of these networks is to provide mechanical stability for cells and tissues. At physiological levels of connectivity, such networks would be mechanically unstable with only central-force interactions. While networks can be stabilized by bending interactions, it has also been shown that they exhibit a critical transition from floppy to rigid as a function of applied strain. Beyond a certain strain threshold, it is predicted that underconstrained networks with only central-force interactions exhibit a discontinuity in the shear modulus. We study the finite-size scaling behavior of this transition and identify both the mechanical discontinuity and critical exponents in the thermodynamic limit. We find both non-mean-field behavior and evidence for a hyperscaling relation for the critical exponents, for which the network stiffness is analogous to the heat capacity for thermal phase transitions. Further evidence for this is also found in the self-averaging properties of fiber networks.
\end{abstract}

\maketitle

\section{Introduction}

\begin{figure}[!h]
	\includegraphics[width=6cm,height=6cm,keepaspectratio]{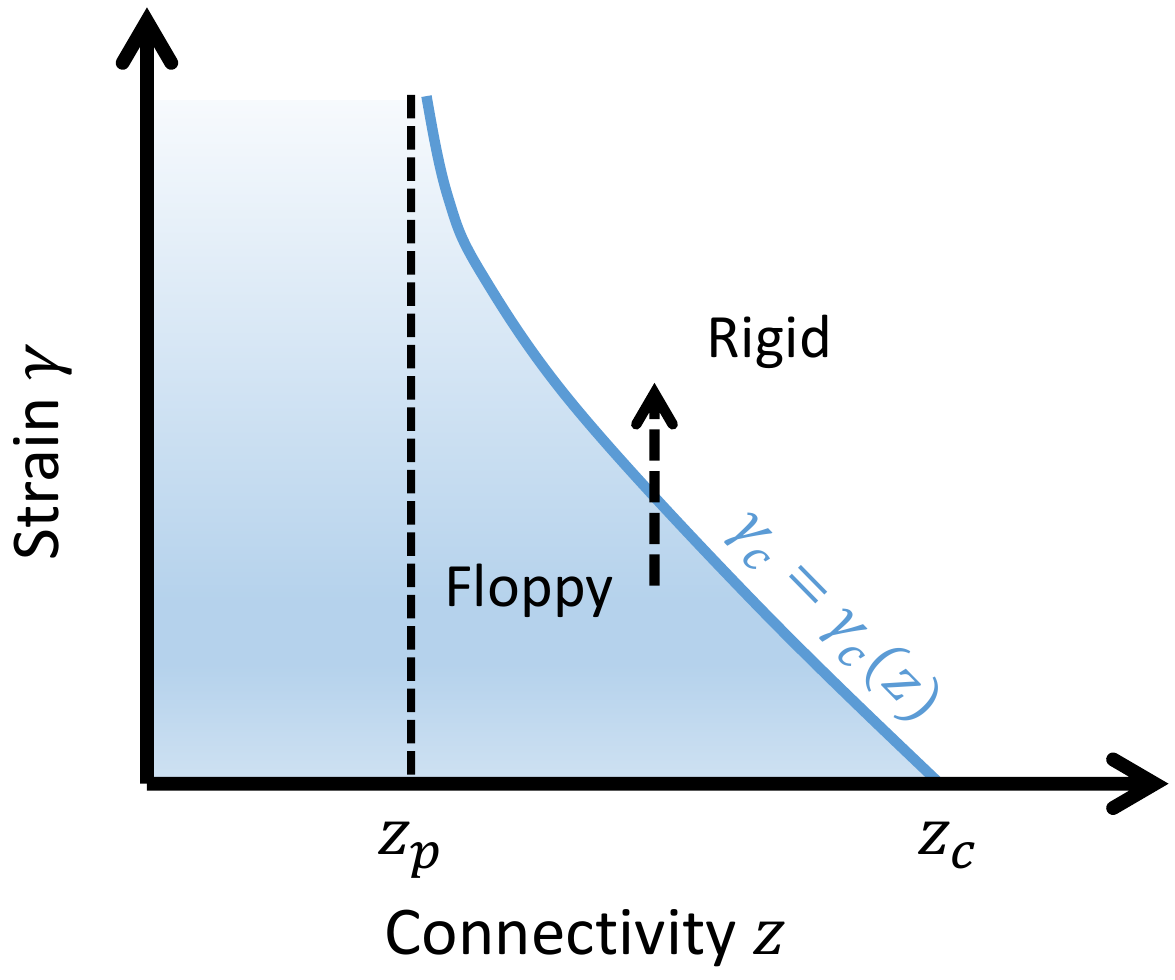}
	\caption{ \label{fig:1} Rigidity phase diagram of central force networks. {Upon increasing the average} connectivity $z$ at $\gamma=0$, {a network} passes through three distinct regimes: (i) a disconnected structure for connectivity less than the percolation connectivity $z<z_p$ (ii) a percolated but floppy network for $z_p<z<z_c \simeq 2d $ and (iii) a rigid network for connectivity greater than $z_c$. Applying a sufficiently large finite strain to an otherwise floppy network with $z_p<z<z_c$ rigidifies the system. For a given $z$ in this range, a critical transition is observed with increasing strain, as indicated by the dashed arrow. The second-order line of transitions is characterized by a critical strain $\gamma_c(z)$ that varies linearly with $z$ near $z_c$ \cite{wyart_elasticity_2008} (see also Fig.\ \ref{fig:A3} in the Appendix).}
\end{figure}

In addition to common thermal phase transitions such as melting or ferromagnetism, there are a number of athermal phase transitions such as rigidity percolation \cite{thorpe_continuous_1983,feng_effective-medium_1985,jacobs_generic_1995} and zero-temperature jamming \cite{cates_jamming_1998,liu_jamming_1998,van_hecke_jamming_2010,bi_jamming_2011,bi_statistical_2015}. These athermal transitions may even exhibit signatures of criticality that are similar to thermal systems. In the case of rigidity percolation, as bond probability or average connectivity $z$ increases on a random central-force network, the number of floppy modes decreases by adding constraints until the isostatic connectivity $z_c$ is reached, at which the system becomes rigid. A simple counting argument by Maxwell shows that $z_c \approx 2d$ where $d$ is dimensionality \cite{maxwell_i.reciprocal_1870,calladine_buckminster_1978}. This linear rigidity transition has been studied in random network models with additional bending interactions \cite{feng_position-space_1985,arbabi_elastic_1988,sahimi_mechanics_1993}. In general, floppy subisostatic central force networks can be stabilized by various mechanisms or additional interactions such as extra springs \cite{wyart_elasticity_2008}, bending resistance \cite{broedersz_criticality_2011}, thermal fluctuations \cite{dennison_fluctuation-stabilized_2013,dennison_critical_2016}, and applied strain \cite{guyon_non-local_1990,sheinman_nonlinear_2012}. 
Sharma et al.\ \cite{sharma_strain-controlled_2016} recently showed that networks with $z<z_c$ exhibit a line of critical floppy-to-rigid transitions under shear deformation and that this line of mechanical phase transitions can account for the nonlinear rheology of collagen networks. The corresponding phase diagram is schematically shown in Fig.\ \ref{fig:1}, where the critical strain $\gamma_c$ at the transition is a function of connectivity $z<z_c$.

\begin{figure}[!h]
	\includegraphics[width=7.5cm,height=7.5cm,keepaspectratio]{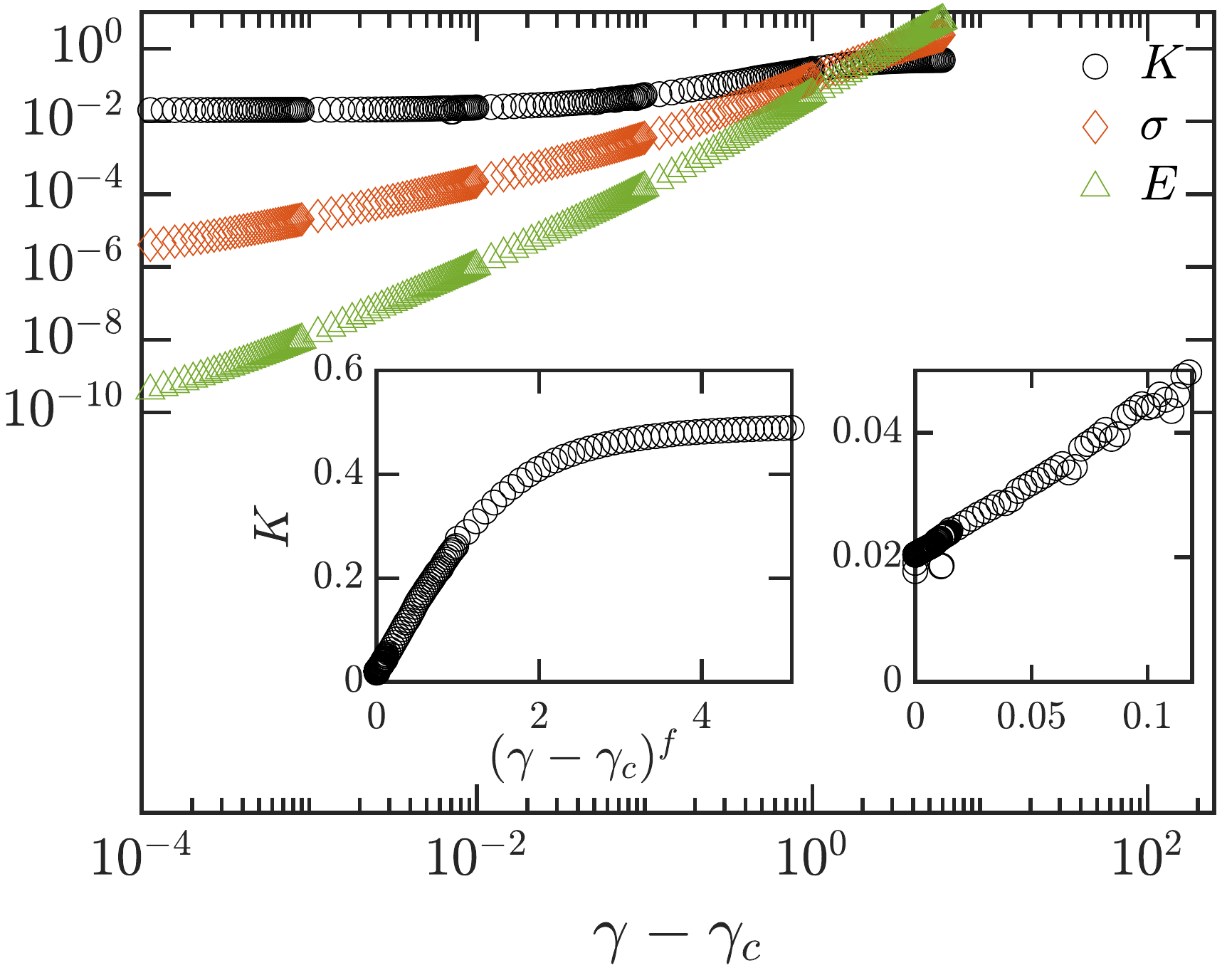}
	\caption{ \label{fig:2} Elastic energy $E$, shear stress $\sigma$, and differential shear modulus $K$ versus excess shear strain to the critical point $\gamma-\gamma_c$ for a single realization of a subisostatic triangular network with $z=3.3$. We use the finite modulus at the critical strain $\gamma_c$ as the shear modulus discontinuity, i.e., $K_c = K(\gamma_c)$. Inset: a linear plot showing the scaling behavior of $K$ for the same sample. By zooming in this plot on the right side, we observe a distinct modulus discontinuity $K_c$.}
\end{figure}

Recent experiments \cite{lindstrom_biopolymer_2010,licup_stress_2015,jansen_role_2018,burla_connectivity_2020} have shown that collagen biopolymers form networks that are in the subisostatic regime with $z<z_c$. It has also been shown that the rheology of such networks is consistent with computational fiber network models that include both strong stretching interactions and weak fiber bending rigidity \cite{licup_stress_2015,sharma_strain-controlled_2016}. Although even a weak bending rigidity tends to suppress the critical signatures of the transition shown in Fig.\ \ref{fig:1}, the critical exponents can still be identified both theoretically and experimentally in a way similar, e.g., to ferromagnetism at non-zero applied field. 
To understand criticality and finite-size effects in the strain-controlled transition, we focus on fiber networks with purely central force interactions as a function of shear strain $\gamma$. At a critical strain $\gamma_c$, there can be a small but finite discontinuity in the differential shear modulus $K=\partial\sigma/\partial\gamma$, where $\sigma$ is the shear stress \cite{vermeulen_geometry_2017,merkel_minimal-length_2019}. Figure \ref{fig:2} shows the macroscopic modulus, shear stress and elastic energy of a diluted triangular network as a function of the distance above its critical strain. Although both elastic energy $E$ and shear stress $\sigma$ approach zero as $\Delta\gamma=\gamma-\gamma_c$ approaches zero from above, the stiffness $K$ exhibits a finite discontinuity $K_c$. The left inset of Fig.\ \ref{fig:2} shows $K$ versus $|\Delta \gamma| ^f$, where $f \neq 1$ is a non-mean-field scaling exponent. The observed straight line in this linear plot illustrates the critical scaling behavior of $K$ near $\gamma_c$. Moreover, a distinct discontinuity in the modulus can be seen in the right inset of Fig.\ \ref{fig:2}, showing the region closer to $\gamma_c$. The scaling behavior of $K$ and the critical exponent $f$ are more systematically studied in the later sections, where
we study the finite-size scaling of the discontinuity and its effect on the scaling exponents, which have also previously been studied using a complementary approach with the addition of small, non-zero bending rigidity \cite{sharma_strain-controlled_2016}. Using these modified exponents, we test scaling relations recently predicted for fiber networks \cite{shivers_scaling_2019}.

\section{Simulation method}

To investigate the stiffness discontinuity in fiber networks, we use various network models including (i) triangular, (ii) \textit{phantomized} triangular \cite{broedersz_criticality_2011,licup_stress_2015}, (iii) 2D and (iv) 3D jammed-packing-derived \cite{wyart_elasticity_2008,tighe_force_2010,baumgarten_normal_2018, merkel_minimal-length_2019,shivers_normal_2019}, (v) Mikado \cite{wilhelm_elasticity_2003,head_deformation_2003}, and (vi) 2D Voronoi network \cite{heussinger_stiff_2006,arzash_stress-stabilized_2019}. Triangular networks are built by depositing individual fibers of length $W$ on a periodic triangular lattice. The lattice spacing is $\ell_0=1$. A full triangular network has an average connectivity of $z=6$. In order to avoid the trivial effects of system-spanning fibers, we initially cut a single random bond from every fiber. Since the number of connections for a crosslink in real biopolymer networks is either 3 (branching point) or 4 (fiber crossing), we enforce this local connectivity in phantomized triangular model. A single node in a full triangular network has three crossing fibers. We \emph{phantomize} the network by detaching one of these fibers randomly for every node \cite{broedersz_molecular_2011,licup_stress_2015}. Therefore, a fully phantomized triangular network has an average connectivity of $z=4$. Similar to the triangular network model, a random bond is removed from every fiber to avoid system-spanning fibers. 

We note that our lattice models are not generic, i.e., the nodes are not displaced from an initial regular lattice. Although generic lattices can be important for linear elasticity \cite{jacobs_generic_1995,moukarzel_elastic_2012}, the nonlinear elasticity studied here is insensitive to small displacements in the the initial configuration, as shown in Ref. \cite{rens_nonlinear_2016}. This is due to the fact that the transition we study occurs at a finite strain threshold, by which significant nonaffine deformation has occurred. 2D (3D) packing-derived networks are generated by randomly placing $N=W^2 \; (W^3)$ disks (spheres) in a periodic box (cube) of length $W$. We use $50/50$ bidisperse particle mixture with radii ratio of $1.4$. These frictionless particles interact via a harmonic soft repulsive potential \cite{ohern_random_2002,ohern_jamming_2003,goodrich_jamming_2014}. The particles are uniformly expanded until the system exhibits both non-zero bulk and shear moduli, i.e., the system is jammed at which a contact network excluding rattlers is derived. This contact network shows an average connectivity of $z \simeq z_c$. Mikado networks are constructed by populating a box of size $W$ with $N$ fibers of length $L$. Permanent crosslinks are introduced at the crossing points between two fibers. Because of the preparation procedure for the Mikado model, the average connectivity of the network approaches $4$ from below as number of fibers $N$ increases. To construct Mikado networks, we choose a line density of $NL^2/W^2 \simeq 7$ that results in an average connectivity of $z \simeq 3.4$. The 2D Voronoi model is prepared by performing a Voronoi tessellation of $W^2/2$ random seeds in a periodic box with side length of $W$, using the CGAL library \cite{the_cgal_project_cgal_2019}. A full Voronoi network has an average connectivity of $z=3$. 

For all network models, we randomly cut bonds until the desired average connectivity $z<z_c$ is reached. Any remaining dangling bonds are removed since they do not contribute to the network's stiffness. The random dilution process not only yields a subisostatic network similar to real biopolymers but also introduces disorder in the system. All crosslinks in our computational models are permanent and freely hinged. An example image of each model is shown in Fig.\ \ref{fig:A1} in the Appendix. Among these computational models, we note that the bond length distribution of Mikado and Voronoi models is similar to the observed filament length distribution of collagen networks \cite{lindstrom_biopolymer_2010}.

In the above models, the bonds are treated as simple Hookean springs. Therefore, the elastic energy of the network is calculated as
\begin{equation}\label{eq:1}
E = \frac{\mu}{2} \sum_{ij}^{}\frac{(\ell_{ij} - \ell_{ij,0})^2}{\ell_{ij,0}},
\end{equation}
in which $\mu$ (in units of energy/length) is the stretching (Young's) modulus of individual bonds, $\ell_{ij}$ and $\ell_{ij,0}$ are the current and rest bond length between nodes $i$ and $j$ respectively. We note that the rest lengths are defined as bond lengths after constructing the networks, i.e., prior to any deformation. The sum is taken over all bonds in the network. We set $\mu=1$ in our simulations.

We apply simple volume-preserving shear deformations in a step-wise procedure with small step size. The deformation tensors in 2D and 3D are as follow
\begin{equation}\label{eq:2}
\Lambda_{2\textrm{D}}(\gamma) = \begin{bmatrix} 
1 & \gamma \\
0 & 1
\end{bmatrix}, \;
\Lambda_{3\textrm{D}}(\gamma) = \begin{bmatrix} 
1 & 0 & \gamma \\
0 & 1 & 0 \\
0 & 0 & 1
\end{bmatrix}
\end{equation}
where $\gamma$ is the shear strain and the networks are sheared in $x$-direction. Note that the 3D networks are deformed in $x-z$ plane.

We assume a quasi-static process, i.e., the system reaches mechanical equilibrium after each deformation step. Therefore, after each strain step, we minimize the elastic energy in Eq.\ \ref{eq:1} using one of the multidimensional minimization algorithms such as FIRE \cite{bitzek_structural_2006}, conjugate gradient \cite{press_numerical_1992}, and BFGS2 method from the GSL library \cite{galassi_et_al_gnu_2018}. To reduce finite size effects, we utilize periodic boundary conditions in both directions. Moreover, we use Lees-Edwards boundary conditions to deform the networks \cite{lees_computer_1972}. After finding the mechanical equilibrium configuration at each strain step, we compute the stress components as follows \cite{shivers_scaling_2019}
\begin{equation}\label{eq:3}
\sigma_{\alpha \beta} = \frac{1}{2V} \sum_{ij}^{}f_{ij,\alpha}r_{ij,\beta},
\end{equation}
in which $V$ is the volume of simulation box, $f_{ij,\alpha}$ is the $\alpha$ component of the force exerted on node $i$ by node $j$, and $r_{ij,\beta}$ is the $\beta$ component of the displacement vector connecting nodes $i$ and $j$. The differential shear modulus $K$ is calculated as $K = d\sigma_{xy}/d\gamma$ in 2D and $K = d\sigma_{xz}/d\gamma$ in 3D at each strain value. To remove any possible asymmetry in $K$, we shear each realization in both positive and negative shear strains. Unless otherwise stated, in order to obtain reliable ensemble averages, we use at least $100$ different realizations for every network model.

\begin{figure}[!h]
	\includegraphics[width=8.5cm,height=8.5cm,keepaspectratio]{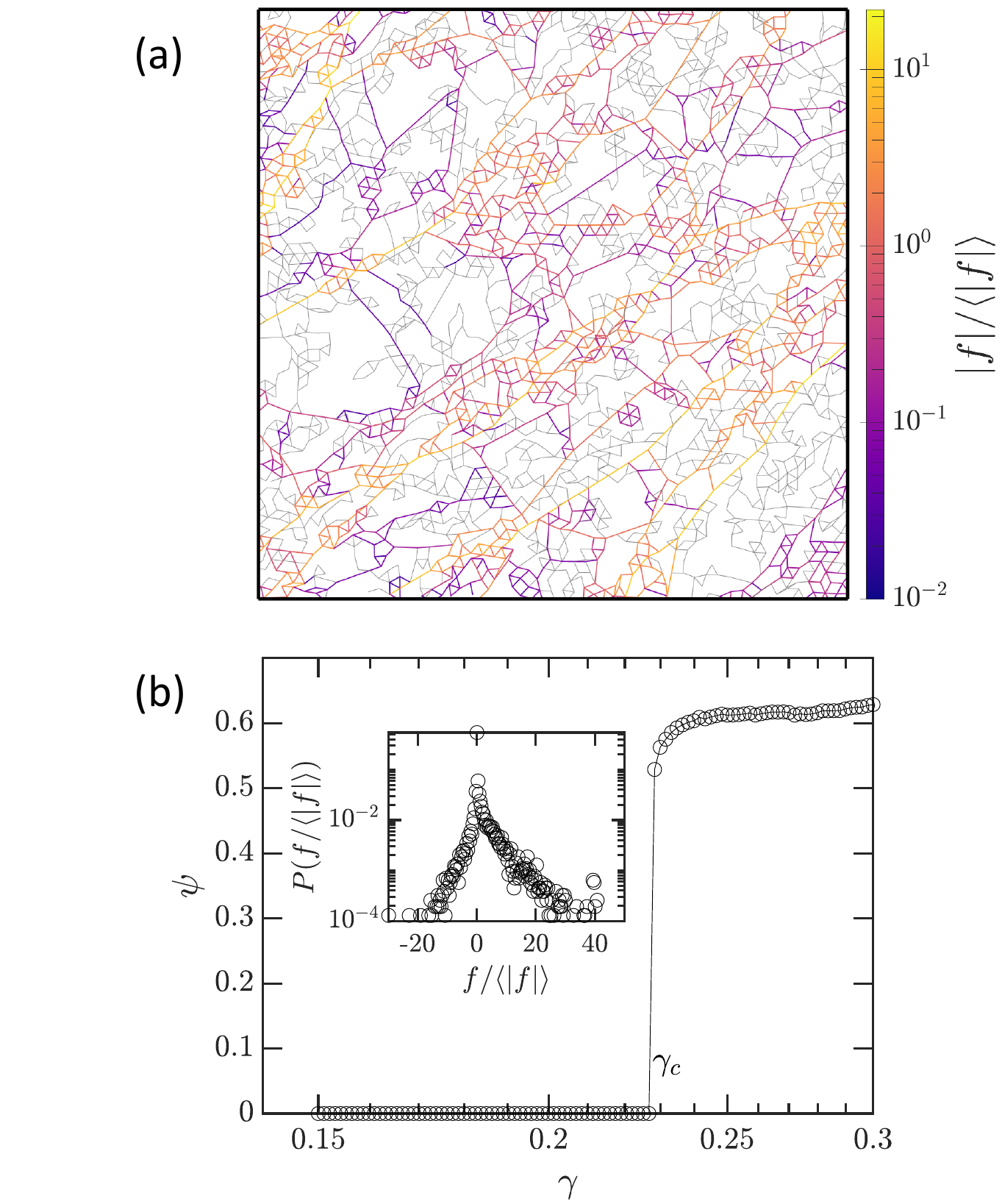}
	\caption{ \label{fig:3} (a) A small section of a triangular network with connectivity $z=3.3$ at the critical strain $\gamma = \gamma_c$. The gray bonds are those with zero force. Bonds with larger forces have a brighter color. This branch-like force chain that appears at the critical strain rigidifies the otherwise floppy network. (b) The participation ratio $\psi$, the ratio of bonds under a finite force to all present bonds, versus shear strain $\gamma$ for the network in (a). As shown, a large portion of bonds undergoes a finite force at the critical strain, i.e., $\psi_c \simeq 0.5$. Inset: the force distributions of the network in (a) at the critical strain, where $\langle |f| \rangle$ is the average of absolute values of bond forces.}
\end{figure}

\section{Results}

By applying shear strain, the subisostatic networks with central force interactions undergo a mechanical phase transition from a floppy to a rigid state \cite{sharma_strain-controlled_2016,sharma_strain-driven_2016}. In contrast to percolation- or jamming-like transitions in which the system rigidifies due to increasing number of bonds or contacts, fiber network models have static structures. Therefore, this floppy-to-rigid transition occurs because of the emergence of finite tension under deformation, here shear strain. The transition point is a function of network's geometry as well as network's connectivity $z$ (see the schematic phase diagram in Fig.\ \ref{fig:1}) . As shown in Fig.\ \ref{fig:3}, a branch-like tensional structure appears at the critical strain that is responsible for the network's rigidity. This rigidity mechanism can be understood in terms of the percolation of these tensional paths. By computing the \textit{participation ratio} $\psi$ as the ratio of bonds with non-zero force to all present bonds in the network, we find that a large portion of the network is under a finite force at the transition point (see Fig.\ \ref{fig:3} b). To calculate $\psi$ we use the absolute value of bond forces $|f_{ij}|$, where $f_{ij}>0$ corresponds to tension. The force distribution at the critical strain is shown in the inset of Fig.\ \ref{fig:3} b. The behavior of this distribution is similar to (compressive) contact force distributions in particle packings \cite{radjai_force_1996,ohern_random_2002,ohern_jamming_2003,wyart_rigidity_2005,majmudar_contact_2005}. Here, however, the distribution shows that there are more tensile than compressive forces at the critical strain, which stabilize the network. Consistent with prior work \cite{shivers_normal_2019}, we find that the force distribution decays exponentially at the critical strain.

To further understand this criticality in central force networks, we investigate the moments of force distribution that are defined as
\begin{equation}\label{}
M_k = \langle \frac{1}{N_b} \sum_{ij} |f_{ij}|^{k} \rangle ,
\end{equation}
in which the angle brackets represent the ensemble average over random realizations, $N_b$ is the number of all bonds, and $|f_{ij}| =|\mu (\ell_{ij} - \ell_{ij,0})/\ell_{ij,0}|$ is the magnitude of force on bond $ij$. Similar to the behavior of percolation on elastic networks \cite{hansen_multifractality_1988,hansen_universality_1989,arbabi_mechanics_1993,sahimi_mechanics_1993}, we find that the moments $M_k$ obey a scaling law near the critical strain 
\begin{equation}\label{}
M_k \sim |\gamma - \gamma_c|^{q_k}.
\end{equation}
This scaling behavior of the first three moments is shown in Fig.\ \ref{fig:A4} in the Appendix. For a triangular network with $z=3.3$, we find that $q_1 = 1.3 \pm 0.1$, $q_2 = 2.5 \pm 0.1$ and $q_3 = 3.7 \pm 0.1$. Interestingly, we observe that $q_k \simeq q_{k-1} +1$ for $k>1$. Note that the zeroth moment of the force distribution is the participation ratio $\psi$ shown in Fig.\ \ref{fig:3}b. The mass fraction of the tensional backbone that appears at the critical strain is given by the participation ratio or zeroth moment at $\gamma_c$ \cite{hansen_universality_1989,bunde_fractals_1995}. In plotting the mass of the tensional structure at the critical strain versus system size $W$, we find that the fractal dimension of this backbone appears to be the same as the euclidean dimension of $2$ (see Fig.\ \ref{fig:A8} in the Appendix).

Of particular interest are the macroscopic properties of fiber networks such as stiffness $K$ near the transition. As we approach the critical point,  we find that $K$ shows a finite discontinuity $K_c$, in agreement with prior work \cite{vermeulen_geometry_2017,merkel_minimal-length_2019}. Figure\ \ref{fig:2} shows the behavior of one random realization of a diluted triangular network very close to its critical strain $\gamma - \gamma_c \simeq 10^{-4}$. In order to find the sample-specific critical point $\gamma_c(W,i)$ for a network with size $W$, we use the bisection method \cite{merkel_minimal-length_2019}. By performing an initial step-wise shearing simulation for every random sample, we first find a strain value $\gamma_{R,i}$ at which the network becomes rigid, i.e., the shear stress calculated from Eq.\ \ref{eq:3} reaches a threshold value. Here we use $10^{-9}$ for the stress threshold. Our results, however, are insensitive to the choice of the threshold value as long as we use a sufficiently small value. The prior strain value to $\gamma_{R,i}$ is considered as the nearest floppy point $\gamma_{F,i}$. Modifying the bracket [$\gamma_{F,i}$, $\gamma_{R,i}$] in at least $20$ bisection steps, we are able to accurately identify the critical point for every random sample $i$. After identifying the critical point, the network is sheared in a step-wise manner from $\gamma_c(W,i)$. Therefore, the final ensemble averages of a specific system size are taken over random realizations with the same distance from their critical strain. Prior work has established that this is a suitable averaging method for finite systems with large disorder \cite{bernardet_disorder_2000}.

As shown previously \cite{vermeulen_geometry_2017} for purely central-force networks, the stiffness $K$ exhibits a scaling behavior with the excess shear strain
\begin{equation}\label{eq:Discont}
K-K_c \sim |\gamma - \gamma_c|^f,
\end{equation}
in which $K_c$ represents a discontinuity in the shear modulus at the transition and $f$ is a non-mean-field exponent.
Subisostatic networks with central force interactions are floppy below this transition. In order to understand the behavior of networks in $\gamma < \gamma_c$ regime, we introduce an additional bending rigidity \cite{broedersz_criticality_2011,licup_stress_2015,shivers_scaling_2019}. In the presence of a weak bending rigidity $\kappa$, the floppy-to-rigid transition in networks becomes a crossover between bend-dominated and stretch-dominated regimes \cite{onck_alternative_2005,sharma_strain-controlled_2016,sharma_strain-driven_2016,shivers_scaling_2019}. In the small strain regime $\gamma < \gamma_c$, the shear modulus is proportional to the bending rigidity $\kappa$ and the following scaling form captures the behavior of $K$ for bend-stabilized fiber networks \cite{sharma_strain-controlled_2016}
\begin{equation}\label{eq:Widom}
K \approx |\gamma - \gamma_c|^f \mathcal{G}_\pm(\kappa/|\gamma - \gamma_c |^\phi),
\end{equation}
in which $\phi$ is a scaling exponent and $\mathcal{G}_\pm$ is the scaling function for regimes above and below the critical strain. In later sections, we discuss in detail the procedure of finding these scaling exponents $f$ and $\phi$.

\begin{figure}[!h]
	\includegraphics[width=8cm,height=8cm,keepaspectratio]{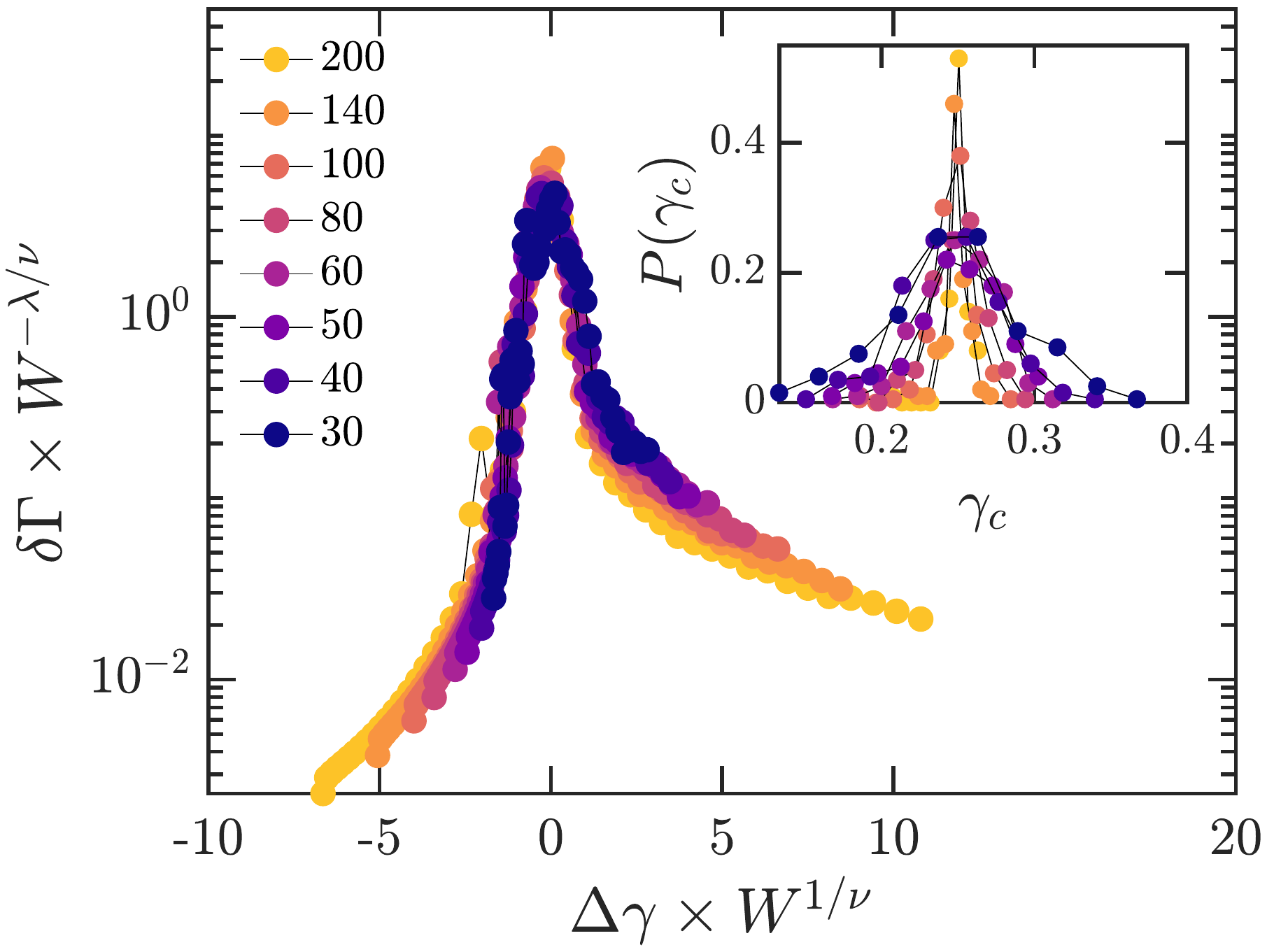}
	\caption{ \label{fig:4} The finite-size collapse of nonaffine fluctuations according to Eq.\ \ref{FluctuationsScaling}. The data are obtained for triangular networks with $z=3.3$ and different lateral size $W$ as specified in the legend. Inset: shows distributions of the critical strain for the same networks.}
\end{figure}

With the scaling exponents $f$ and $\phi$ obtained, we repeat the tests previously carried out for the scaling theory in Ref.\ \cite{shivers_scaling_2019}. Specifically, we consider the finite-size scaling of the nonaffine fluctuations of a diluted triangular network in Fig.\ \ref{fig:4}. The nonaffine displacements are measured by the differential nonaffinity parameter defined as
\begin{equation}\label{eq:4}
\delta \Gamma = \frac{\langle || \delta \mathbf{u}^{\text{NA}} ||^2 \rangle}{\ell^2 \delta \gamma^2},
\end{equation}
in which $\ell$ is the typical bond length of the network, and $\delta \mathbf{u}^{\text{NA}} = \mathbf{u} - \mathbf{u}^{\text{affine}}$ is the nonaffine displacement of a node that is caused by applying an infinitesimal shear strain $\delta \gamma$. To better illustrate this parameter, we show the nonaffine displacement vectors of nodes for a diluted triangular network before, at and after the critical strain in Fig.\ \ref{fig:A5} in the Appendix \cite{sharma_strain-driven_2016}. The differential nonaffinity $\delta \Gamma$ diverges at the critical strain for central force networks, with a susceptibility-like exponent $\lambda = \phi - f$, i.e., $\delta \Gamma \sim |\Delta \gamma|^{-\lambda}$ \cite{broedersz_mechanics_2011,sharma_strain-driven_2016,shivers_scaling_2019}. Moreover, as the system approaches the critical strain, the correlation length diverges as $\xi \sim |\Delta \gamma|^{-\nu}$. When the correlation length is smaller than the system size $W$, i.e., $|\Delta \gamma| \times W^{1/\nu}>1$, we should find $\delta \Gamma \sim |\Delta \gamma|^{-\lambda}$. Near the critical strain, however, the finite-size effects result in $\delta \Gamma \sim |\Delta \gamma|^{\lambda/\nu}$. Therefore, the following scaling form must capture the behavior of fluctuations \cite{sharma_strain-driven_2016} 
\begin{equation}\label{FluctuationsScaling}
\delta \Gamma = W^{\lambda/\nu} \mathcal{H}(\Delta \gamma W^{1/\nu}),
\end{equation}
where the scaling function $\mathcal{H}(x)$ is constant for $|x|<1$ and $|x|^{-\lambda}$ otherwise. The differential nonaffinity is shown for different system sizes of a diluted triangular network in Fig.\ \ref{fig:A5} in the Appendix. Based on the above scaling form, we perform a finite-size scaling analysis as shown in Fig.\ \ref{fig:4}. The correlation length exponent $\nu$ is computed from the hyperscaling relation $f=d\nu - 2$ obtained for this transition in prior work \cite{shivers_scaling_2019}, using the exponent $f$ that is computed by considering the stiffness discontinuity. This excellent collapse of fluctuations further emphasizes the true critical nature of the transition as well as consistency with the hyperscaling relation $f=d\nu - 2$ in fiber networks, even accounting for the discontinuity in $K$. As noted before, this discontinuity has no bearing on the order of the transition, since $K$ is not the order parameter, and is more analogous to the heat capacity in a thermal phase transition \cite{shivers_scaling_2019}. The inset of Fig.\ \ref{fig:4} shows the distribution of critical strains for the same networks in the main figure. As system size increases, the critical strain distribution becomes narrower. Although we focus on finite-size effects in computational fiber models primarily in order to properly identify the behavior of such networks in the thermodynamic limit, we note that experimental rheology on physical collagen networks can also be strongly affected by the sample size, e.g., in sample size dependence of the yield strain \cite{arevalo_size-dependent_2010}. This is likely due to the rather large mesh size of order 10 $\mu$m in many of the experimental studies.

As indicated above, the exponent $f$ is analogous to the heat capacity exponent $\alpha$ in thermal critical phenomena, but with opposite sign. Based on the Harris criterion \cite{harris_effect_1974}, a positive $f>0$ (i.e., $\alpha<0$), for which $\nu > 2/d$, implies that weak randomness does not change the behavior of critical fiber networks. Closely related to the Harris criterion is the \textit{self-averaging} property in critical phenomena. Any observable $X=E$, $\sigma$ or $K$ has different values for different random samples. Therefore for a system with size $W$, we can define for observable $X$ a probability distribution function $P(X,W)$, which is characterized by its average $\langle X \rangle$ and variance $V(X) = \langle X^2 \rangle - \langle X \rangle ^2$. A system is self-averaging if the relative variance $R_V(X) = V(X)/\langle X \rangle^2 \rightarrow 0$ as $W \rightarrow \infty$. In other words, the ensemble average of a self-averaging system does not depend on the disorder introduced by random samples as the system size becomes infinite. 

\begin{figure}[!h]
	\includegraphics[width=12cm,height=12cm,keepaspectratio]{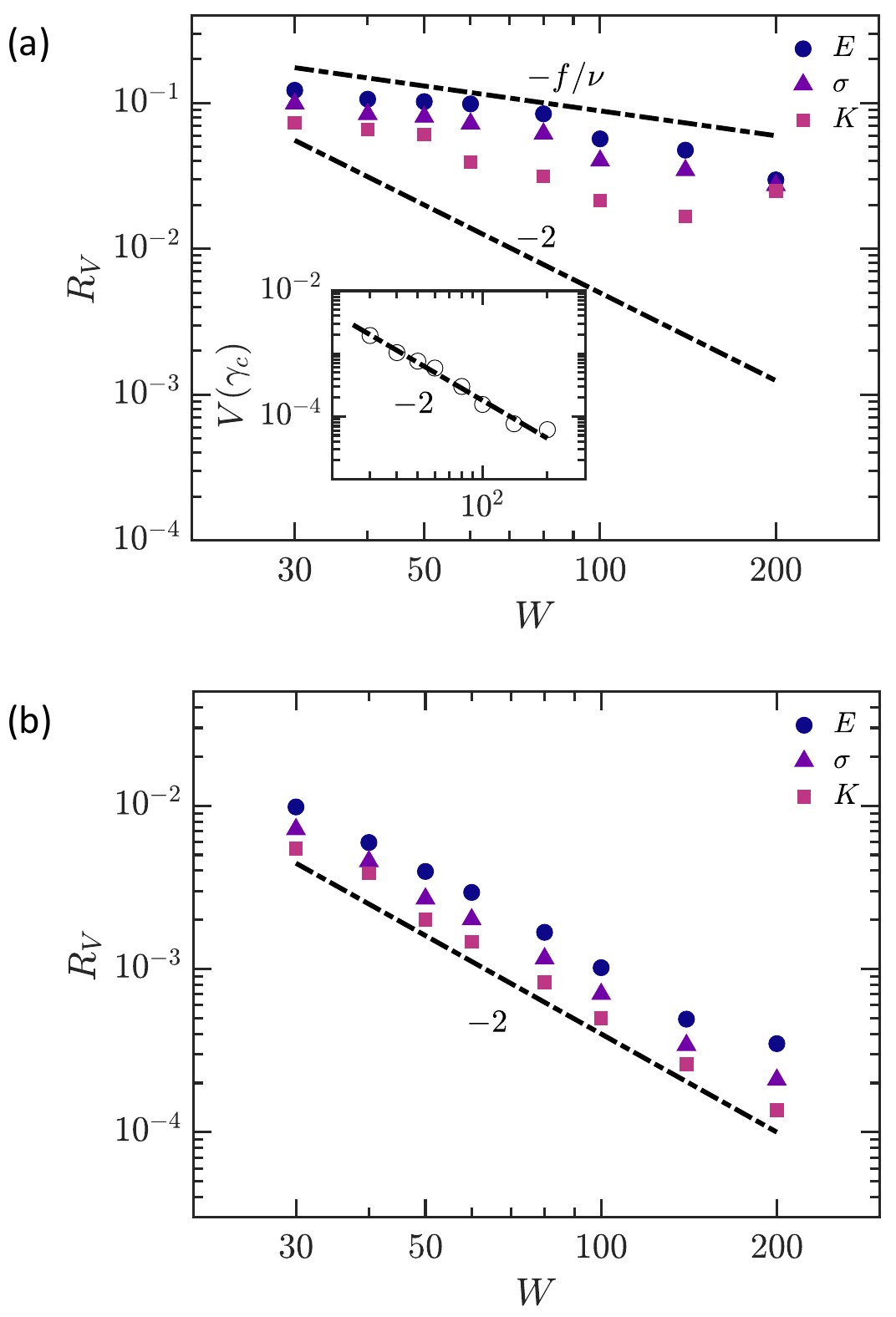}
	\caption{ \label{fig:5} (a) The relative variance of different quantities specified in the legend at the critical strain for a triangular network with $z=3.3$ versus linear system size $W$. Inset: the scaling behavior of variance of critical strains versus system size for the same model. (b) The relative variance of the macroscopic quantities as specified in the legend for the same model in (a) away from the critical strain versus linear system size $W$.}
\end{figure}

Far from the transition, where the system size $W$ is much larger than the correlation length $\xi$, the Brout argument \cite{brout_statistical_1959}, which is based on the central limit theorem, indicates \textit{strong} self-averaging $R_V(X) \sim W^{-d}$ where $d$ is dimensionality \cite{wiseman_lack_1995}. Indeed, for our 2D fiber networks away from the critical strain, we find that the relative variance of macroscopic properties decreases with system size as $W^{-2}$, i.e., fiber networks exhibit strong self-averaging off criticality (see Fig.\ \ref{fig:5}b). Near the transition, however, the correlation length becomes larger than the system size $W \ll \xi$ and the Brout argument does not hold. Therefore, at criticality there is no reason to expect $R_V(X) \sim W^{-d}$ \cite{wiseman_lack_1995,aharony_absence_1996,wiseman_self-averaging_1998}. For example, it is established that $R_V(X)$ shows a $W$-independent behavior, i.e., no self-averaging at the percolation transition for the mass of spanning cluster \cite{stauffer_introduction_2003} and the conductance of diluted resistor networks \cite{harris_randomly_1987}. A \textit{weak} self-averaging, that corresponds to $R_V(X) \sim W^{-a}$ with $0<a<d$, has been identified in bond-diluted Ashkin-Teller models \cite{wiseman_lack_1995}. As proved by Aharony and Harris \cite{aharony_absence_1996}, when randomness is irrelevant, i.e., $\nu > 2/d$ the system exhibits a weak self-averaging behavior where $R_X \sim W^{\alpha/\nu}$ (in our fiber networks $R_X \sim W^{-f/\nu}$). As shown in Fig.\ \ref{fig:5} a, fiber networks appear to exhibit a weak self-averaging at the critical strain, with an exponent close to $f/\nu$. We note that $R_V(X)$ in Fig.\ \ref{fig:5} a is computed in the regime where $|\Delta \gamma| \times W^{1/\nu} \approx 1$. We also find that the variance of critical strains decreases as $V(\gamma_c) \sim W^{-2}$ (see the inset of Fig.\ \ref{fig:5} a), in accordance with Aharony and Harris prediction \cite{aharony_absence_1996}.

\begin{figure}[!h]
	\includegraphics[width=8cm,height=8cm,keepaspectratio]{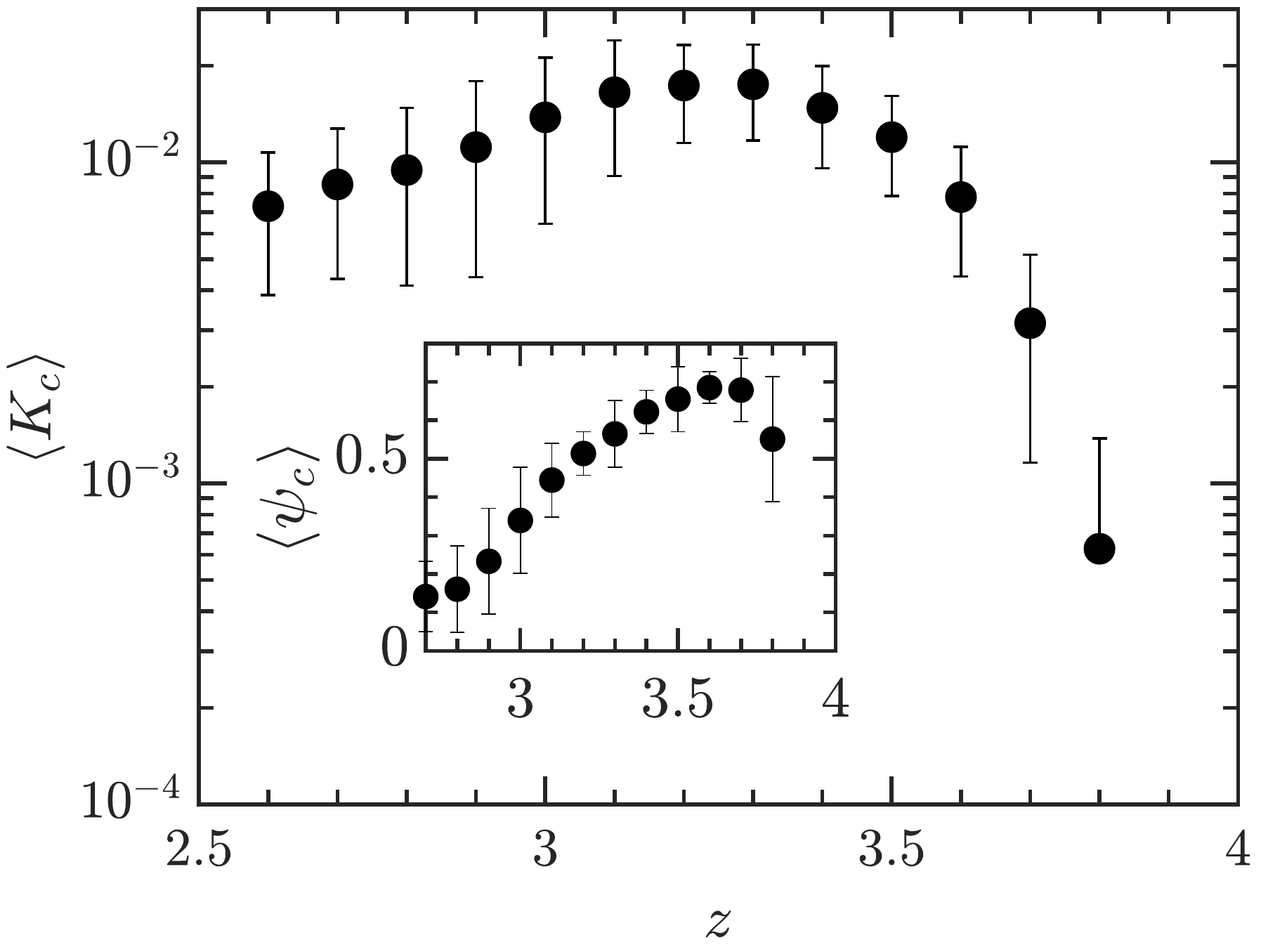}
	\caption{ \label{fig:6} Shear modulus discontinuity versus connectivity $z$ for a triangular network. As connectivity $z$ approaches the isostatic point $z_c$, the jump in shear modulus vanishes $K_c \rightarrow 0$. On the other hand, for networks with low connectivity, a small tensional pattern is responsible for the rigidity of the system. Therefore, $K_c$ decreases as $z$ decreases towards the percolation connectivity. Inset: participation ratio at the critical strain versus connectivity $z$.}
\end{figure}

As prior work showed \cite{wyart_elasticity_2008,merkel_minimal-length_2019}, the shear modulus discontinuity $K_c$ vanishes as network connectivity $z$ approaches the isostatic threshold $z_c=2d$. Figure\ \ref{fig:6} shows the behavior of $K_c$ versus network connectivity $z$. As expected, $K_c$ decreases as $z$ approaches $z_c$. Moreover, as $z$ decreases towards the connectivity percolation transition for a randomly diluted triangular network, we observe a decreasing trend in $K_c$. This regime can be explained by plotting the participation ratio at the critical strain $\psi_c$ in the inset of Fig.\ \ref{fig:6}. As we see $\psi_c$ has a small value for networks with $z$ close to the percolation connectivity. These small tensional patterns are responsible for the network's rigidity at critical strain, hence resulting in lower modulus discontinuity $K_c$.

\begin{figure}[!h]
	\includegraphics[width=12cm,height=12cm,keepaspectratio]{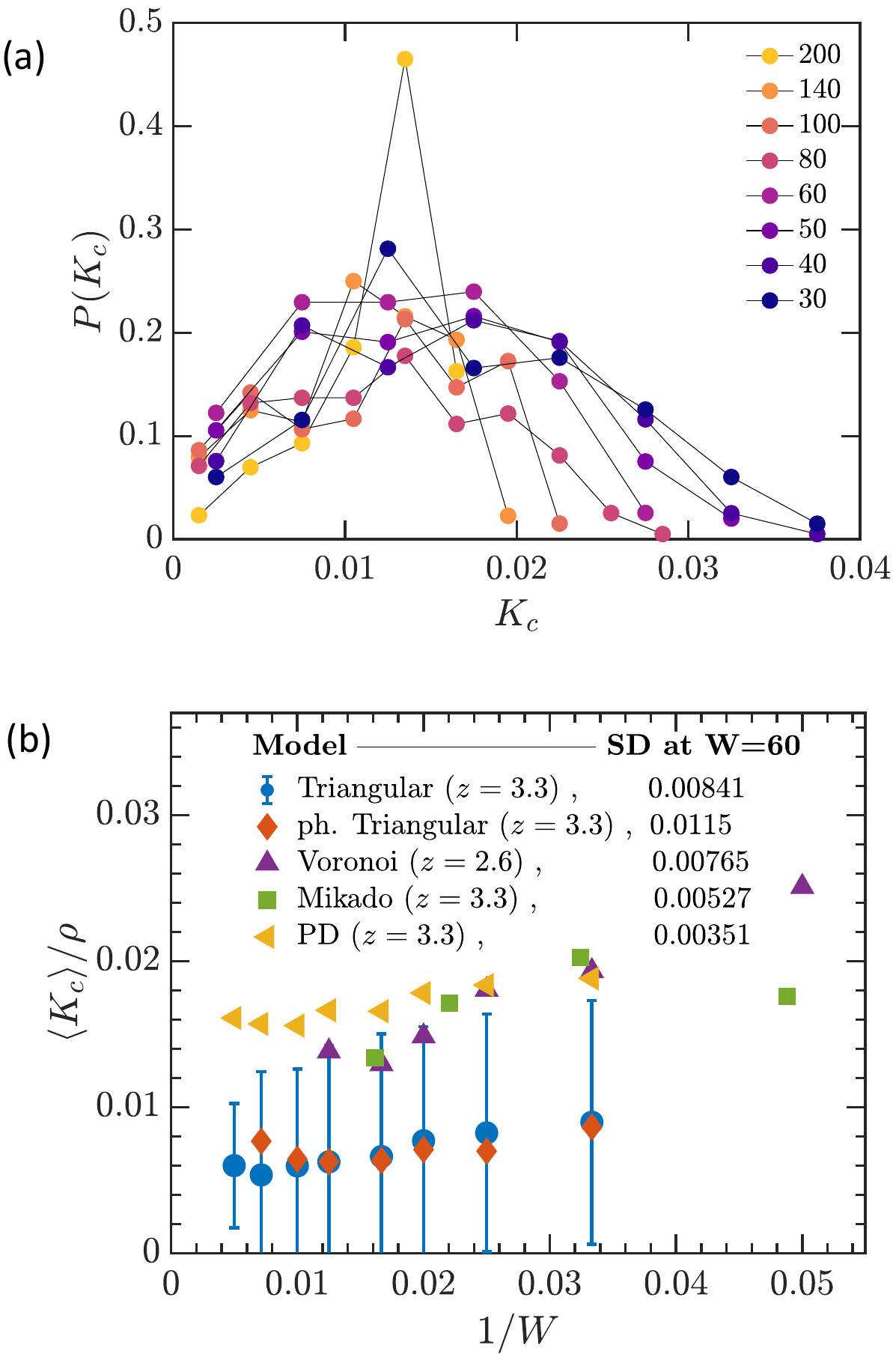}
	\caption{ \label{fig:7} (a) The distributions of shear modulus discontinuity $K_c$ for triangular networks with $z=3.3$ and different system sizes as specified in the legend. (b) Shear modulus discontinuity $K_c$ versus inverse system size $1/W$, for various 2D network models as specified in the legend (For Mikado model we used square root of present nodes in the network as $W$). The data are normalized with the length density $\rho$ for every model. The standard deviations are only shown for the triangular network, though the standard deviation at $W=60$ for every model is shown in the legend.}
\end{figure}

In order to understand the network behavior in the thermodynamic limit, we study the finite-size effects in more detail. One trivial finite-size effect is observed by studying the participation ratio $\psi$. For small number of random realizations, a strand-like percolated force chain, which appears at the critical strain, continues to bear tensions under deformation. This effect results in a plateau in network stiffness $K$, as shown in Fig.\ \ref{fig:A7} in the Appendix. This plateau effect is more prevalent in network models with long, straight fibers such as the triangular model. We next explore the finite-size effects of stiffness discontinuity in fiber networks. The distributions of $K_c$ for various system size are shown in Fig.\ \ref{fig:7} a. The mean of these distributions versus inverse system size exhibits a slow decreasing trend for all different network models (Fig.\ \ref{fig:7} b). However, we find that this discontinuity remains finite but small (of order $0.01$) for all network models as we approach the thermodynamic limit $1/W \rightarrow 0$, consistent with findings of Ref. \cite{vermeulen_geometry_2017} for the Mikado model. This is similar to the behavior of the linear bulk modulus for sphere packings at the jamming transition, which exhibits a finite discontinuity in $z$ in the thermodynamic limit \cite{wyart_rigidity_2005,moukarzel_elastic_2012,goodrich_finite-size_2012,goodrich_scaling_2016}. Vermeulen et al. \cite{vermeulen_geometry_2017} argued that the nonlinear shear modulus discontinuity in fiber networks is due to an emerging single state of self-stress at the network's critical strain. Consistent with this, we find a non-fractal stress backbone at the critical strain.

As mentioned above, the stiffness exponent $f$ has a non-mean-field value, i.e., $f \neq 1$. In fiber networks, the correlation length scales as $\xi \sim \Delta \gamma^{-\nu}$. True critical behavior in simulation results such as ours should only be apparent when the correlation length remains smaller than the system size, i.e., $|\Delta \gamma| \times W^{1/\nu} >1$ \cite{sharma_strain-controlled_2016,shivers_scaling_2019}. Near the critical point, however, the correlation length diverges and the stiffness scales as $K-K_c \sim W^{-f/\nu}$. Therefore, the following scaling function captures the stiffness behavior
\begin{equation}
K-K_c  = W^{-f/\nu} \mathcal{F}(\Delta \gamma W^{1/\nu}),
\end{equation}
in which the function $\mathcal{F}(x)$ is a constant for $x<1$ and $x^f$ for $x>1$. Note that we are only able to investigate one side of the transition $\Delta \gamma >0$ for central force networks.
 
 \begin{figure}[!h]
 	\includegraphics[width=12cm,height=12cm,keepaspectratio]{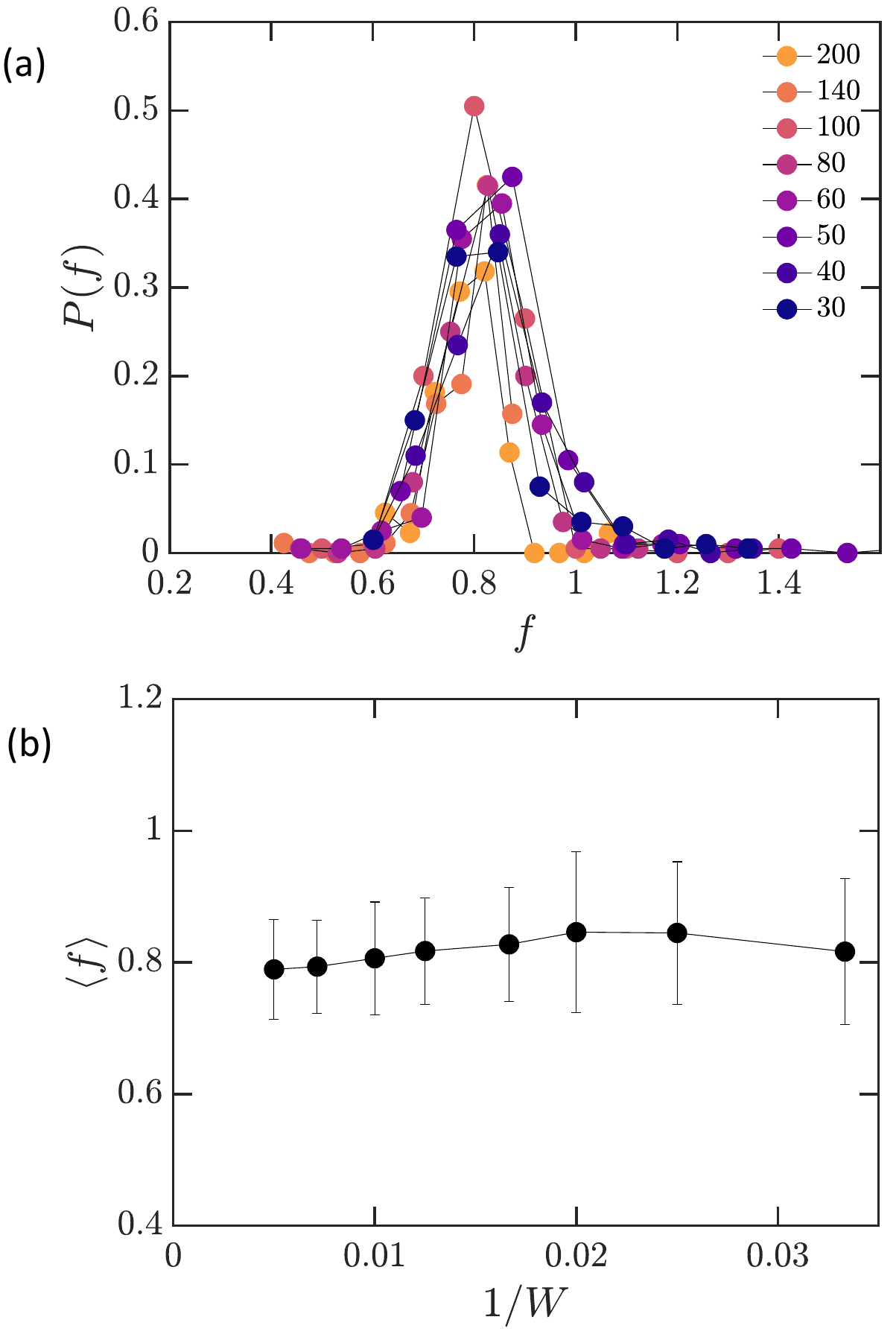}
 	\caption{ \label{fig:8} (a) The distributions of the stiffness exponents $f$ for different system sizes for a triangular network with $z=3.3$. The exponents are obtained in the critical regime in which $|\Delta \gamma| \times W^{1/\nu} > 1.0$ for all sizes. (b) The ensemble average of $f$, which is obtained from the distributions in (a), versus inverse system size $1/W$. The error bars are showing the standard deviations of samples.}
 \end{figure}
 
To obtain the stiffness exponent $f$, we implement a power-law fit of $K - K_c$ versus $\gamma - \gamma_c$ for every individual sample of different system sizes in the critical regime, where $|\Delta \gamma| \times W^{1/\nu} >1$ for every size $W$. We use sample-dependent $K_c$ and $\gamma_c$. Figure\ \ref{fig:8} a shows the $f$ distributions for different system sizes for a triangular network with $z=3.3$. The average of these distributions are shown in Fig.\ \ref{fig:8} b. As can be observed, we find negligible differences in $f$ for different system sizes when the exponents are obtained in the true critical regime. However, instead of this size-dependent approach, if the scaling exponents $f$ are collected in a fixed strain window for all sizes, a size-dependent behavior of $f$ is unavoidable due to the finite-size effects (see Fig.\ \ref{fig:A9} in the Appendix). We conclude an $f = 0.79 \pm 0.07$ corresponding to $W=140$ for triangular networks with $z=3.3$.
 
\begin{figure}[!h]
 	\includegraphics[width=12cm,height=12cm,keepaspectratio]{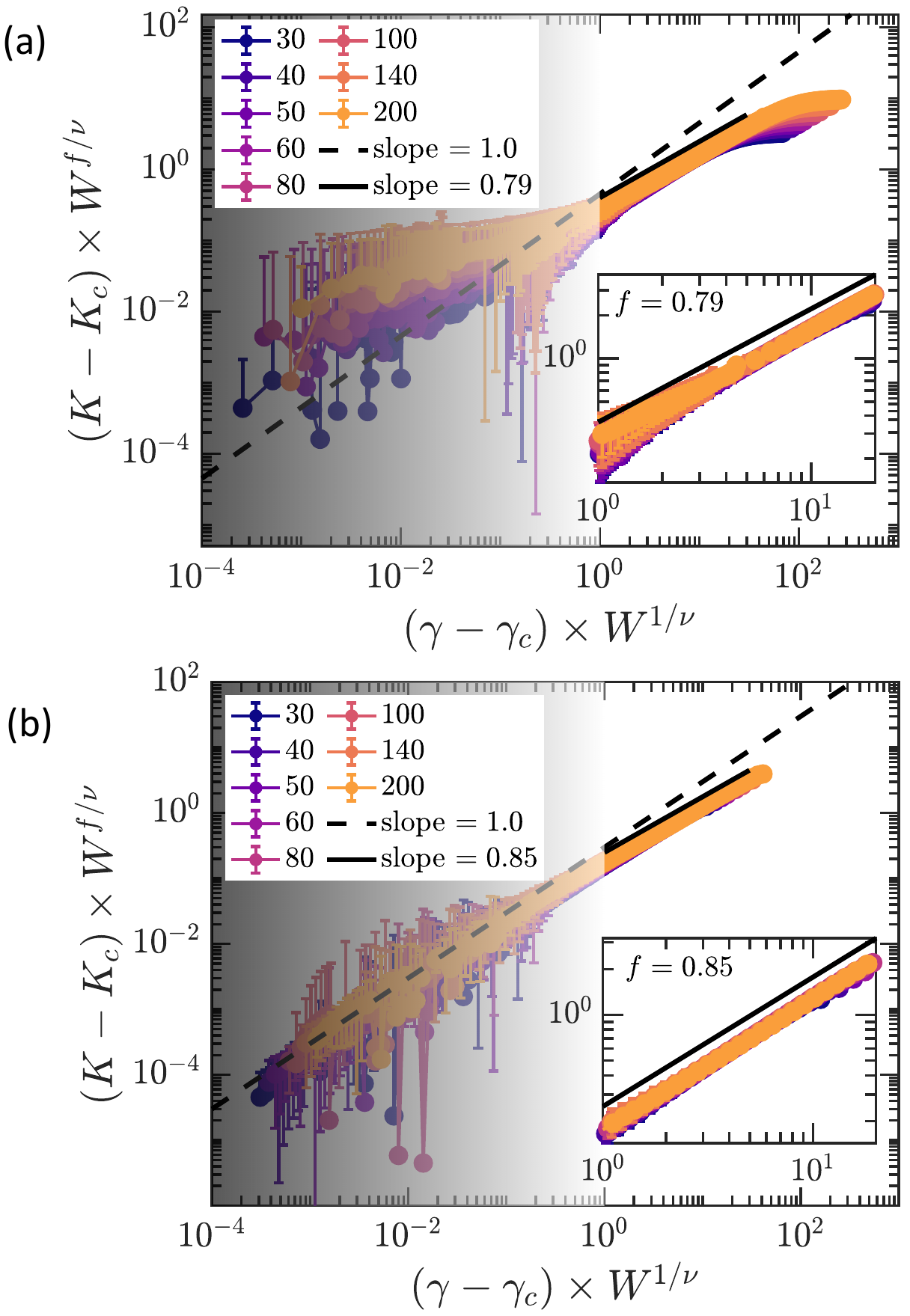}
 	\caption{ \label{fig:9} (a) Finite-size scaling of $K-K_c$ for a triangular network with $z=3.3$. The inset shows the collapse of data in the critical regime with $f=0.79 \pm 0.07$. (b) A similar finite-size scaling as in (a) for a 2D jammed-packing-derived model with $z=3.3$. A distinct analytic regime, i.e., a slope of $1.0$ can be observed in this model as $\gamma - \gamma_c \rightarrow 0$. The inset, however, shows the non-mean-field exponent $f=0.85 \pm 0.05$ in the critical regime. The finite-size dominated regime is shaded in both plots.}
\end{figure}
 
By performing an extensive finite-size scaling analysis of the stiffness data for the diluted triangular model in Fig.\ \ref{fig:9} a, we find three distinct regimes: (i) a finite-size dominated region for $|\Delta \gamma| \times W^{1/\nu} \lesssim1.0$, (ii) a true critical regime for $1\lesssim |\Delta \gamma| \times W^{1/\nu}$ and (iii) an eventual large strain regime outside of the critical regime. By using the hyperscaling relation $f = d\nu -2$, $f$ is the only remaining free parameter used for the analysis in Fig.\ \ref{fig:9} a.  As shown in the inset of Fig.\ \ref{fig:9} a, we are able to collapse the data in the critical regime by using $f=0.79 \pm 0.07$ for a randomly diluted triangular network with $z=3.3$. 
A similar finite-size scaling analysis performed for randomly diluted, 2D jammed-packing-derived networks with $z=3.3$ in Fig.\ \ref{fig:9} b results in a consistent exponent $f=0.85 \pm 0.05$. In agreement with computational studies in 3D \cite{sharma_strain-controlled_2016,sharma_strain-driven_2016}, we also find a non-mean-field $f <1.0$ for 3D jammed-packing-derived networks with $z=3.3$ (see Fig.\ \ref{fig:A10} in the Appendix). This exponent, however, is obtained using only one system size $W=20$. Further work will be needed for a detailed finite-size scaling analysis in 3D similar to Fig.\ \ref{fig:9}. Nevertheless, prior work has shown a high degree of consistency between the 2D and (the somewhat more limited) 3D simulations. Moreover, experiments on collagen networks have so far shown consistency with 2D models \cite{sharma_strain-controlled_2016,jansen_role_2018}. Thus, we have good reason to believe that our conclusions are not limited to idealized 2D systems.

We note that the exponents we observe are robust to changes or errors in the value of the discontinuity $K_c$ in the critical regime (ii) (see Fig.\ \ref{fig:A11} in the Appendix). By performing the same analysis in Fig.\ \ref{fig:9} a, for instance, but using the modulus discontinuity in the thermodynamic limit $K_c^\infty$ instead of sample-dependent $K_c$, we obtain the same scaling exponent $f$, provided that $|\Delta \gamma| \times W^{1/\nu}\gtrsim1$ (see Fig.\ \ref{fig:A12} in the Appendix).
Thus, we limit our analysis of the critical exponents to the regime (ii) with $|\Delta \gamma| \times W^{1/\nu}\gtrsim1$, where we find consistent values of $f\simeq0.79-0.85$, as also reported for Mikado networks previously in Ref.\ \cite{vermeulen_geometry_2017}.
These results are, however, inconsistent with Ref.\ \cite{merkel_minimal-length_2019}, where it was argued that $f=1$ should be generic for fiber networks. We note that it is possible to observe an apparent $f=1$ regime due to finite size effects, as we clearly observe in Fig.\ \ref{fig:9} b when the system size is smaller than of order $|\Delta\gamma|^{-\nu}$.
The apparent exponent $f$ in this case, however, would then not be a critical exponent \cite{stauffer_introduction_2003,binder_monte_2010}.  A natural explanation for an apparent exponent of $1.0$ here can simply be the first term in a scaling function that becomes analytic (and not critical) for a finite system, as has been argued for packings of soft, frictionless particles \cite{goodrich_finite-size_2012}. We note that the finite-size scaling analysis studied here is a rather general technique for understanding critical phenomena in finite-size computer simulations. Hence, we expect that a similar approach in thermal gel models with intermolecular interactions \cite{peleg_filamentous_2007,kroger_formation_2008,peleg_effect_2009} will provide insights about their critical phase transition.

As mentioned before, the sub-isostatic central-force networks can be stabilized by adding bending resistance to fibers. Figure.\ \ref{fig:A13} a in the Appendix shows the shear modulus versus strain for diluted triangular networks with different bending rigidity $\kappa$. For such bend-stabilized networks, the shear modulus is captured by the scaling form of Eq.\ \ref{eq:Widom}. To find the exponent $\phi$ in Eq.\ \ref{eq:Widom}, we fit a power-law to the stiffness data in the regime where $\gamma < \gamma_c$, in which we have $K \approx \kappa |\gamma - \gamma_c|^{f-\phi}$. For individual samples, we find $\phi$ using the corresponding $f$ exponents that are already collected for central-force networks. For a triangular network with $z=3.3$, we find $\phi = 2.64 \pm 0.12$ that is obtained by using system size $W=100$ and $\kappa = 10^{-5}$. The inset of Fig.\ \ref{fig:A13} b in the Appendix shows the distribution of $\phi$. Using these values of $f$ and $\phi$, a Widom-like scaling collapse corresponding to Eq.\ \ref{eq:Widom} is shown in Fig.\ \ref{fig:A13} b and c in the Appendix, for individual samples and the ensemble average of data respectively.

\section{Summary and Discussion}

In this work, we focus on the critical signatures of mechanical phase transitions in central-force fiber networks as a function of shear strain. As the applied strain approaches a critical value $\gamma_c$ from above, the stress is borne by a sparse, branch-like structure that is responsible for network stability. By analyzing various moments of the force distributions, we identify scaling exponents for these moments near the transition, similar to prior work on rigidity percolation \cite{hansen_multifractality_1988,hansen_universality_1989,arbabi_mechanics_1993,sahimi_mechanics_1993}. We also find that the fractal dimension of the load-bearing structure at the critical strain appears to be $2.0$ in 2D. This is consistent with a finite value of the participation ratio $\psi$, as well as a finite discontinuity in the network stiffness $K$ in the thermodynamic limit $W\rightarrow\infty$. 

Further, we study the self-averaging properties of this athermal critical phase transition. We observe a strong self-averaging off criticality, i.e., with relative variance $R_V(X) \sim W^{-d}$ for $X = E$, $\sigma$ and $K$. This is consistent with what is expected for thermal systems, based on the Brout argument \cite{brout_statistical_1959}. At criticality, however, as the correlation length $\xi$ reaches or becomes larger than the system size $W$, we find a weak self-averaging of all macroscopic properties $E$, $\sigma$, and $K$ at the critical strain. Specifically, $R_V(X) \sim W^{-a}$ with $0<a<d$. This weak self-averaging at the critical point is in agreement with thermal systems that satisfy the Harris criterion \cite{harris_effect_1974}, i.e., for which the heat capacity exponent $\alpha<0$. As argued in Ref.\ \cite{shivers_scaling_2019}, the network stiffness is analogous to heat capacity but with the stiffness exponent $f=-\alpha$. Thus, our observations of weak self-averaging provide further evidence for this analogy and suggest that the mechanical critical behavior along the line of transitions in Fig.\ \ref{fig:1} should be insensitive to weak disorder. 

By simulating various network models, we confirm that fiber networks exhibit a finite shear modulus discontinuity $K_c$, in agreement with Refs.\ \cite{vermeulen_geometry_2017,merkel_minimal-length_2019}.  We observe a weakly decreasing trend in $K_c$ as a function of system size, but with a non-zero value in the thermodynamic limit. This discontinuity does, however, vanish as the network connectivity $z$ approaches the isostatic point $z_c$, consistent with Refs.\ \cite{wyart_elasticity_2008,merkel_minimal-length_2019}. We also find that this discontinuity decreases as one approaches connectivity percolation. 
We show that allowing for this discontinuity slightly modifies the scaling exponents obtained previously for fiber networks using other methods. The discrepancies between these methods, however, are within the estimated error bars. 

Moreover, by repeating the finite-size scaling analysis of the nonaffine fluctuations from Ref.\ \cite{shivers_scaling_2019} we again find evidence for the hyperscaling relation $f=d\nu -2$ \cite{shivers_scaling_2019} and non-mean-field nature of the transition. In estimating the stiffness exponent $f$, we perform an extensive finite-size scaling analysis that reveals three distinct regimes; besides a critical region with non-mean-field exponents, we find a finite-size dominated region for $|\Delta \gamma| \times W^{1/\nu} <1.0$, as well as an off critical regime for large strains. In the finite-size dominated regime, we show that the stiffness exponent may appear to be consistent with the mean-field value $f=1$ (Fig.\ 9). As noted above, however, this may simply be due to analyticity for finite systems and may have no bearing on possible mean-field behavior. This may explain some reports of mean-field behavior, such as in Ref.\ \cite{merkel_minimal-length_2019}. It is important to emphasize that the scaling exponents cannot be reliably extracted from simulations close to the transition, i.e., for small $|\Delta\gamma|\rightarrow0$, where $|\Delta \gamma| \times W^{1/\nu} \lesssim1$.

\section*{Acknowledgments}

This work was supported in part by the National Science Foundation Division of Materials Research (Grant DMR1826623) and the National Science Foundation Center for Theoretical Biological Physics (Grant PHY-1427654). J.L.S. acknowledges the support of the Riki Kobayashi Fellowship in Chemical Engineering and the Ken Kennedy Institute Oil \& Gas HPC Conference Fellowship. We also acknowledge useful conversations with Andrea Liu, Tom Lubensky and Lisa Manning.  

%

\newpage
\onecolumngrid
\appendix
\renewcommand\thefigure{A.\arabic{figure}} 
\setcounter{figure}{0}
\renewcommand\appendixname{}
\renewcommand{\theequation}{A.\arabic{equation}}
\setcounter{equation}{0}
 
 \section*{Appendix}
 
\subsection*{Network models}

\begin{figure}[!h]
	\includegraphics[width=16cm,height=16cm,keepaspectratio]{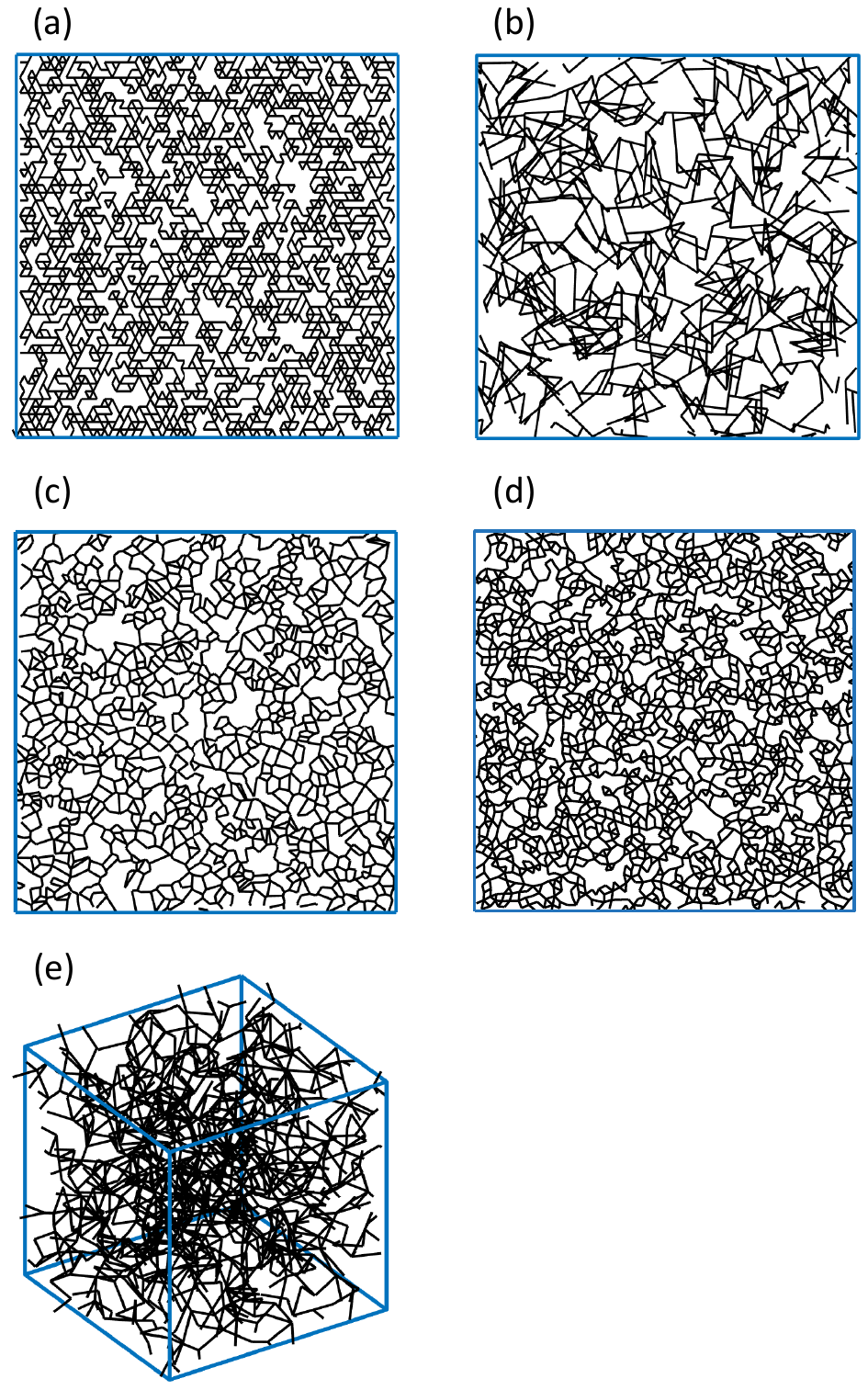}
	\caption{ \label{fig:A1} Snapshots of various network models. (a) Randomly diluted triangular network with $z=3.3$. (b) Randomly diluted Mikado model with $z=3.3$. (c) Randomly diluted 2D Voronoi network with $z=2.6$. (d) Randomly diluted 2D jammed-packing-derived network with $z=3.3$. (e) Randomly diluted 3D jammed-packing-derived network with $z=3.3$.}
\end{figure}
\FloatBarrier

\begin{figure}[!h]
	\includegraphics[width=8.5cm,height=8.5cm,keepaspectratio]{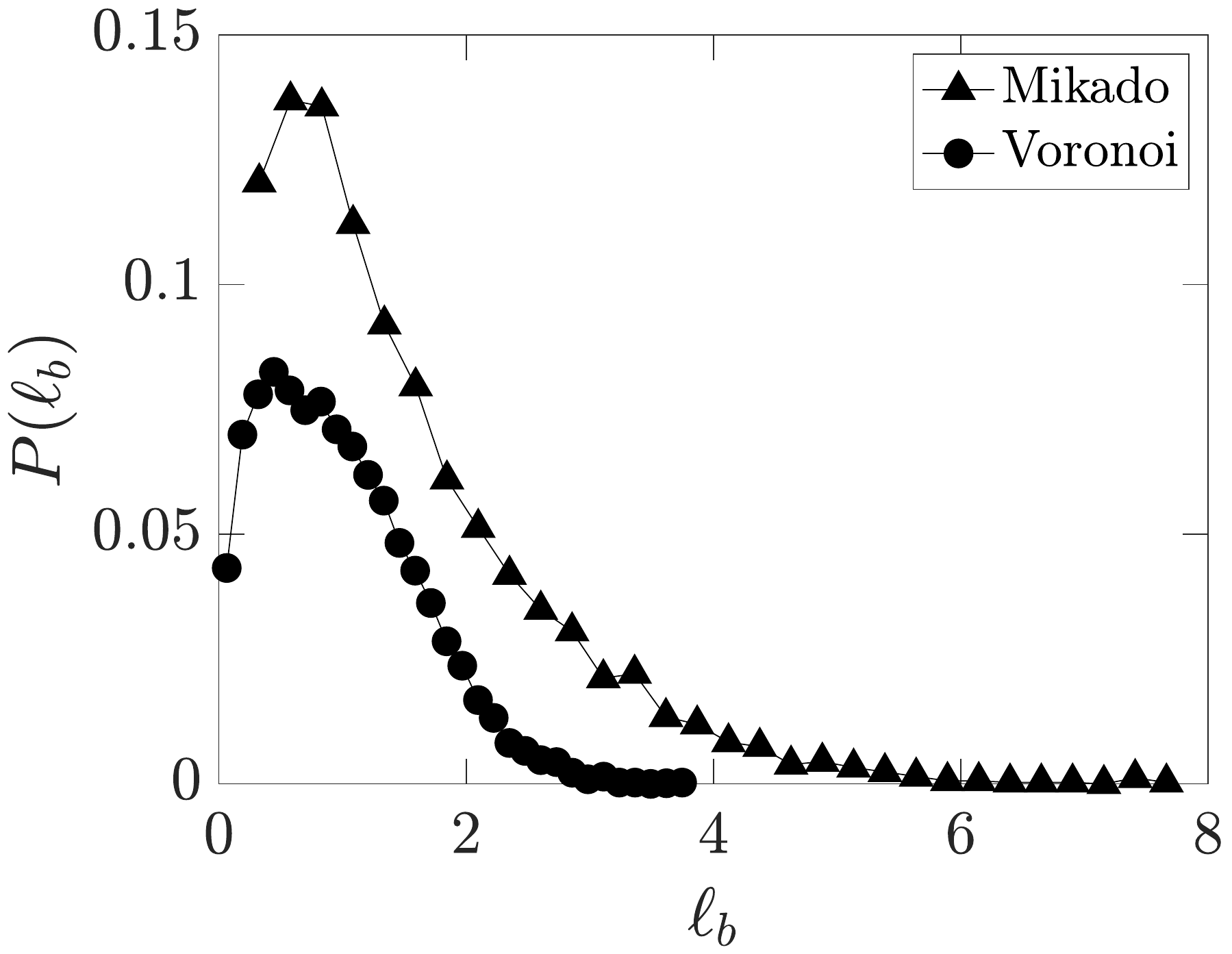}
	\caption{ \label{fig:A2} The bond length distribution of Mikado and Voronoi models. These exponential-like decay of bond length has been identified in real collagen networks.}
\end{figure}
\FloatBarrier

\begin{figure}[!h]
	\includegraphics[width=8.5cm,height=8.5cm,keepaspectratio]{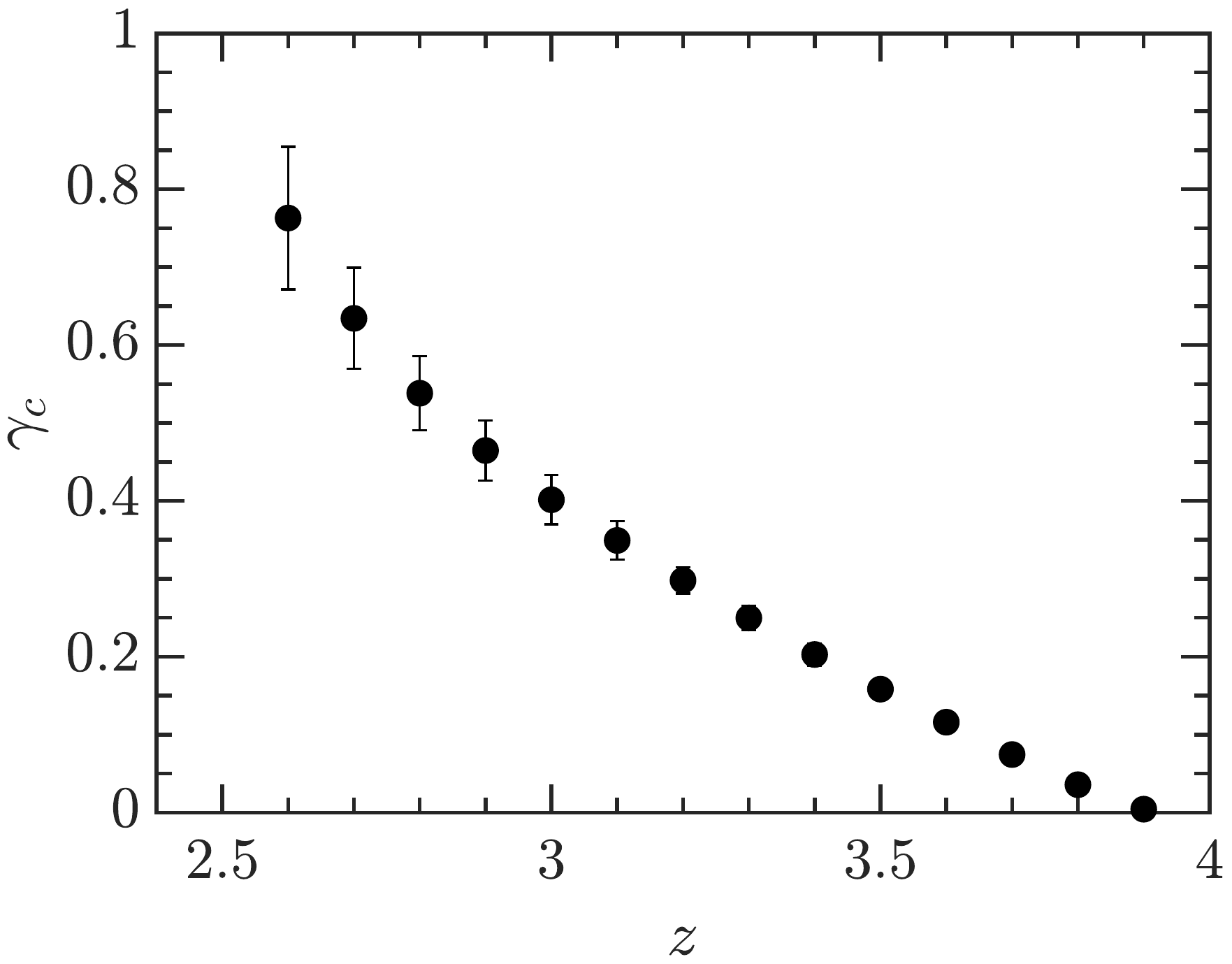}
	\caption{ \label{fig:A3} The critical strain versus connectivity for a randomly diluted triangular network with size $W=80$. Near the isostatic point $z_c$, the relation appears to be linear. Note that $z_c <4.0$ is due to the finite size effects.}
\end{figure}
\FloatBarrier

\subsection*{Scaling of the moments of force distributions}

\begin{figure}[!h]
	\includegraphics[width=8.5cm,height=8.5cm,keepaspectratio]{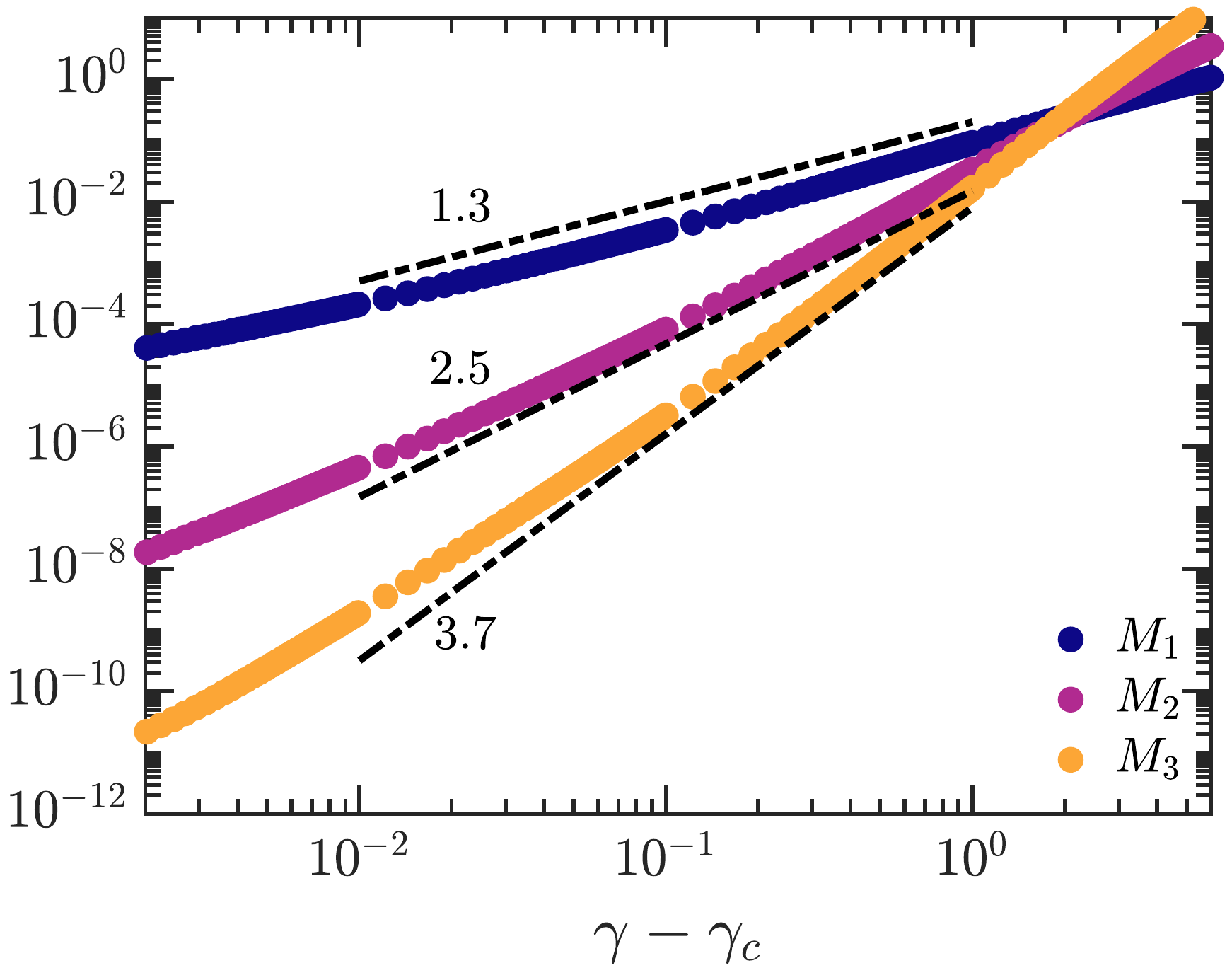}
	\caption{ \label{fig:A4} The scaling behavior of first three moments of force distribution versus excess strain to the critical point for a triangular network with $z=3.3$.}
\end{figure}
\FloatBarrier

\subsection*{Nonaffine displacement fluctuations}

In order to find the correlation length exponent $\nu$, we compute the nonaffine fluctuations in athermal fiber networks. The differential nonaffinity parameter $\delta \Gamma$ defined in Eq.\ 7 measures the nonaffine node displacements after applying a small shear strain from a previous state.
\begin{figure}[!h]
	\includegraphics[width=12cm,height=12cm,keepaspectratio]{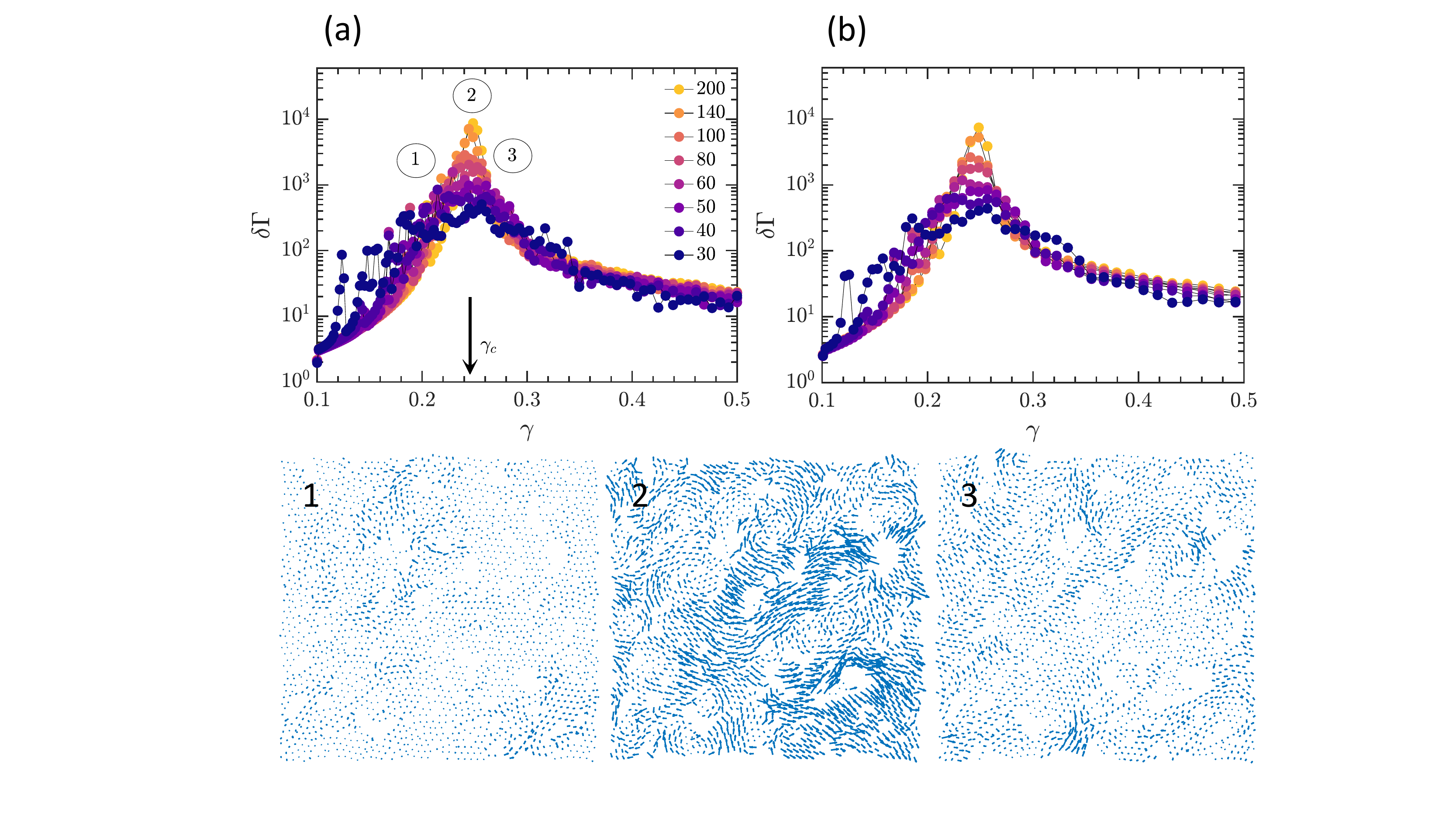}
	\caption{ \label{fig:A5} (a) The unscaled differential nonaffinity parameter defined in Eq.\ 7 in the main text for diluted triangular networks with $z=3.3$ and sizes as shown in the legend. The nonaffine displacement vectors of a single sample of size $W=50$ are shown for a strain value less than (1) at (2) and greater than (3) the critical strain $\gamma_c$. (b) Coarse-grained $\delta \Gamma$, using local averaging of every two adjacent data points in (a).}
\end{figure}
\FloatBarrier
Figure\ \ref{fig:A5} a shows the differential nonaffinity for diluted triangular network with $z=3.3$ for different system sizes. The nonaffine vectors of network's nodes for a single sample of size $W=50$ are shown at (1): $\gamma < \gamma_c$ (2): $\gamma = \gamma_c$ (3): $\gamma > \gamma_c$. As we can see from the displacement field, large nonaffine node displacements are evident at the critical strain, which corresponds to the peak in differential nonaffinity parameter. In order to reduce the noise in $\delta \Gamma$ for finite-size scaling, we use the local averaging method; every two adjacent values of Fig.\ \ref{fig:A5} a are averaged and the result is shown in Fig.\ \ref{fig:A5} b. The finite-size collapse shown in Fig.\ 4 in the main text is indeed the collapse of coarse-grained data in Fig.\ \ref{fig:A5} b.

\subsection*{Finite size analysis of the participation ratio $\psi$}

\begin{figure}[!h]
	\includegraphics[width=8.5cm,height=8.5cm,keepaspectratio]{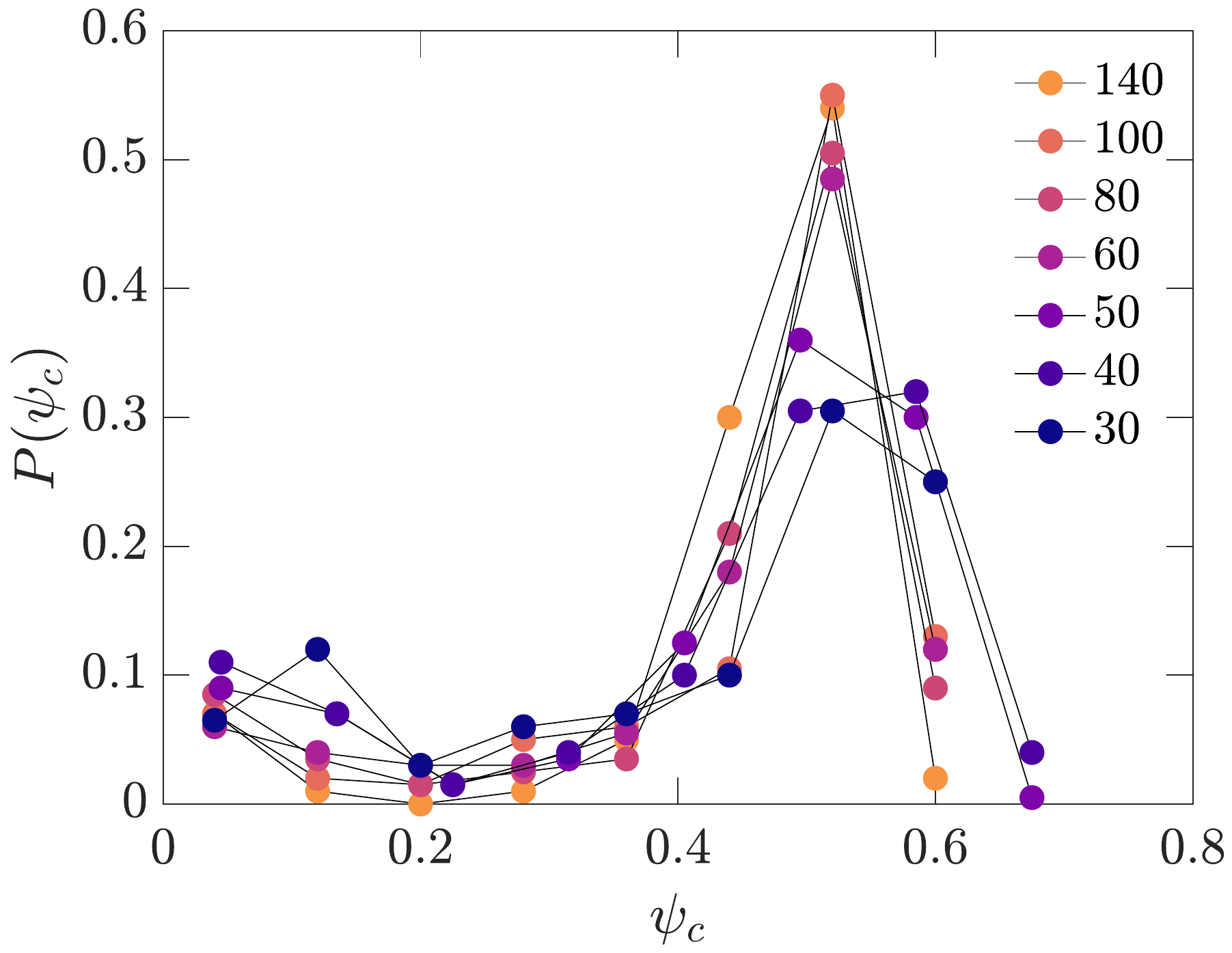}
	\caption{ \label{fig:A6} The distributions of critical participation ratio $\psi_c$ for different sizes of a triangular network with $z=3.3$.}
\end{figure}

\begin{figure}[!h]
	\includegraphics[width=10cm,height=10cm,keepaspectratio]{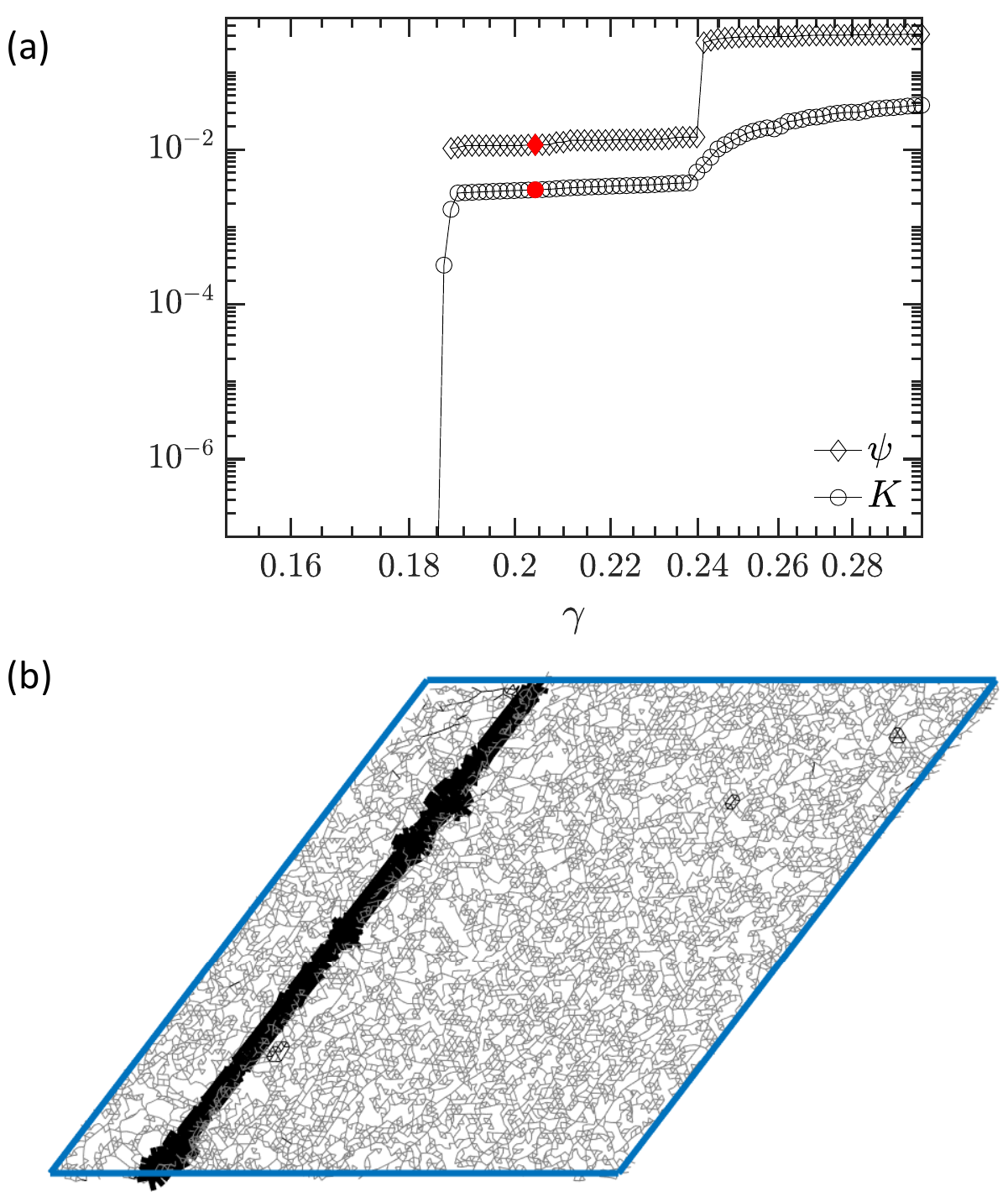}
	\caption{ \label{fig:A7} (a) The participation ratio $\psi$ and stiffness $K$ for a single random realizations with a plateau effect for diluted triangular model with $z=3.3$ and $W=100$. (b) The tensional line responsible for the plateau effect near the critical strain in (a) is shown by plotting bonds with a thickness proportional to their tensions at the highlighted strain point in (a).}
\end{figure}

The distribution of participation ratio at the critical strain $\psi_c$ is shown in Fig.\ \ref{fig:A6} for diluted triangular networks at various sizes. The distribution appears to be bimodal: the large peak is related to the branch-like force chains in the network, similar to the structure shown in Fig.\ 3 a, and the small peak at low participation ratio, which is due to the finite-size effects. Although the location of large peak depends on the network connectivity $z$, the small peak is the result of a small number of realizations that shows a tensional path at the critical strain connecting upper and lower sides of the periodic box. This tension line yields a plateau behavior in stiffness of the network (see Fig.\ \ref{fig:A7} a). As system size increases, the number of samples with this small tensional structure decreases and disappears completely in the thermodynamic limit. This tensional pattern is shown for a single sample in Fig.\ \ref{fig:A7} b.

\FloatBarrier

\FloatBarrier

\begin{figure}[!h]
	\includegraphics[width=8.5cm,height=8.5cm,keepaspectratio]{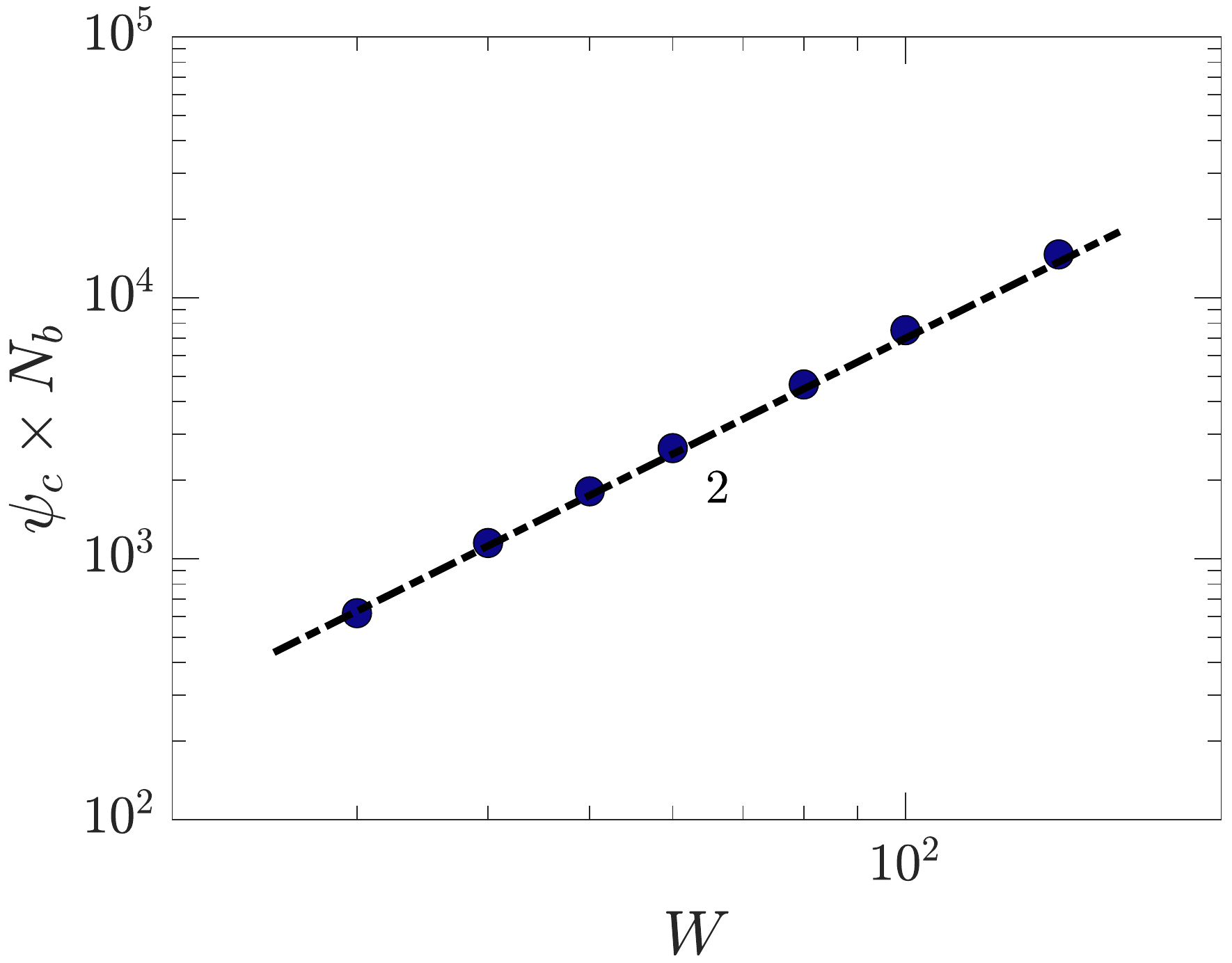}
	\caption{ \label{fig:A8} The critical participation ratio times the number of bonds, which is a measure of mass of the tensional structure at the critical point, versus network size for a triangular model with $z=3.3$.}
\end{figure}
\FloatBarrier

\subsection*{Finite size effects on the scaling exponent $f$}

\begin{figure}[!h]
	\includegraphics[width=8.5cm,height=8.5cm,keepaspectratio]{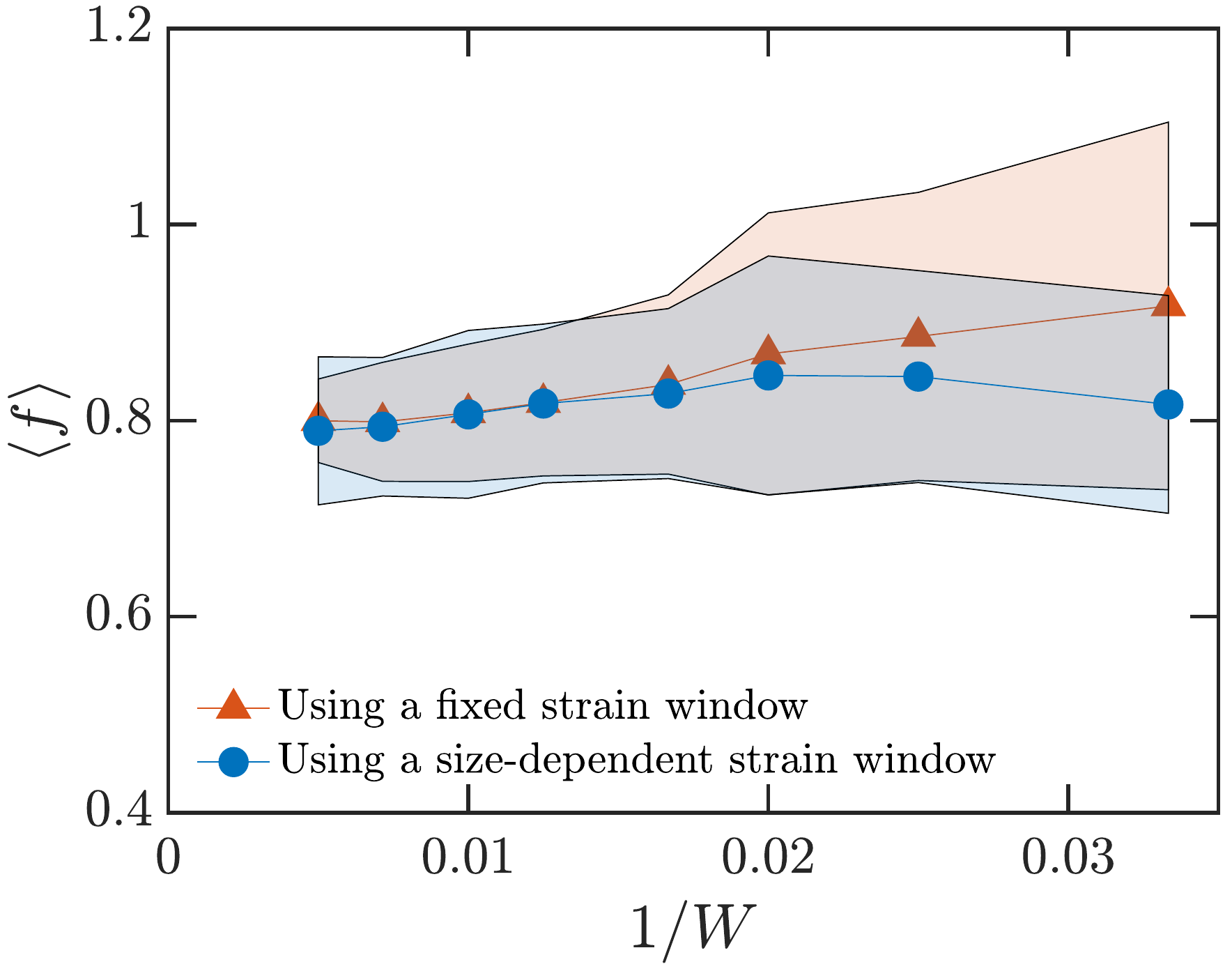}
	\caption{ \label{fig:A9} Comparing two methods of finding $f$ for different sizes of a triangular network with $z=3.3$. The shadow area is showing the standard deviations. The red triangles correspond to the exponents that are obtained in a fixed strain window for all sizes, here the strain window is $\Delta \gamma = 0.055 - 1.0$. The blue circles correspond to the exponents we obtained in a size-dependent strain window in which $1.0 <|\Delta \gamma| \times W^{1/\nu} < 30$ for all sizes.}
\end{figure}
\FloatBarrier

\subsection*{$f$ exponent for a 3D network}

We obtain $f = 0.84 \pm 0.13$ for 3D jammed-packing-derived model with $z=3.3$. The data are collected for only one system size $W=20$, averaging over $40$ random samples. Assuming the hyperscaling relation $f = d\nu -2$ holds in 3D, we used $\nu = (f + 2)/3 \approxeq 0.95$ for the following scaling plot. This network has $\gamma_c = 0.57 \pm 0.03$ and $K_c = 0.006 \pm 0.004$.
Future studies will be needed in 3D for a detailed finite-size scaling analysis similar to Fig.\ 9 in the main text as well as testing the hyperscaling relation $f = d\nu -2$.

\begin{figure}[!h]
	\includegraphics[width=8.5cm,height=8.5cm,keepaspectratio]{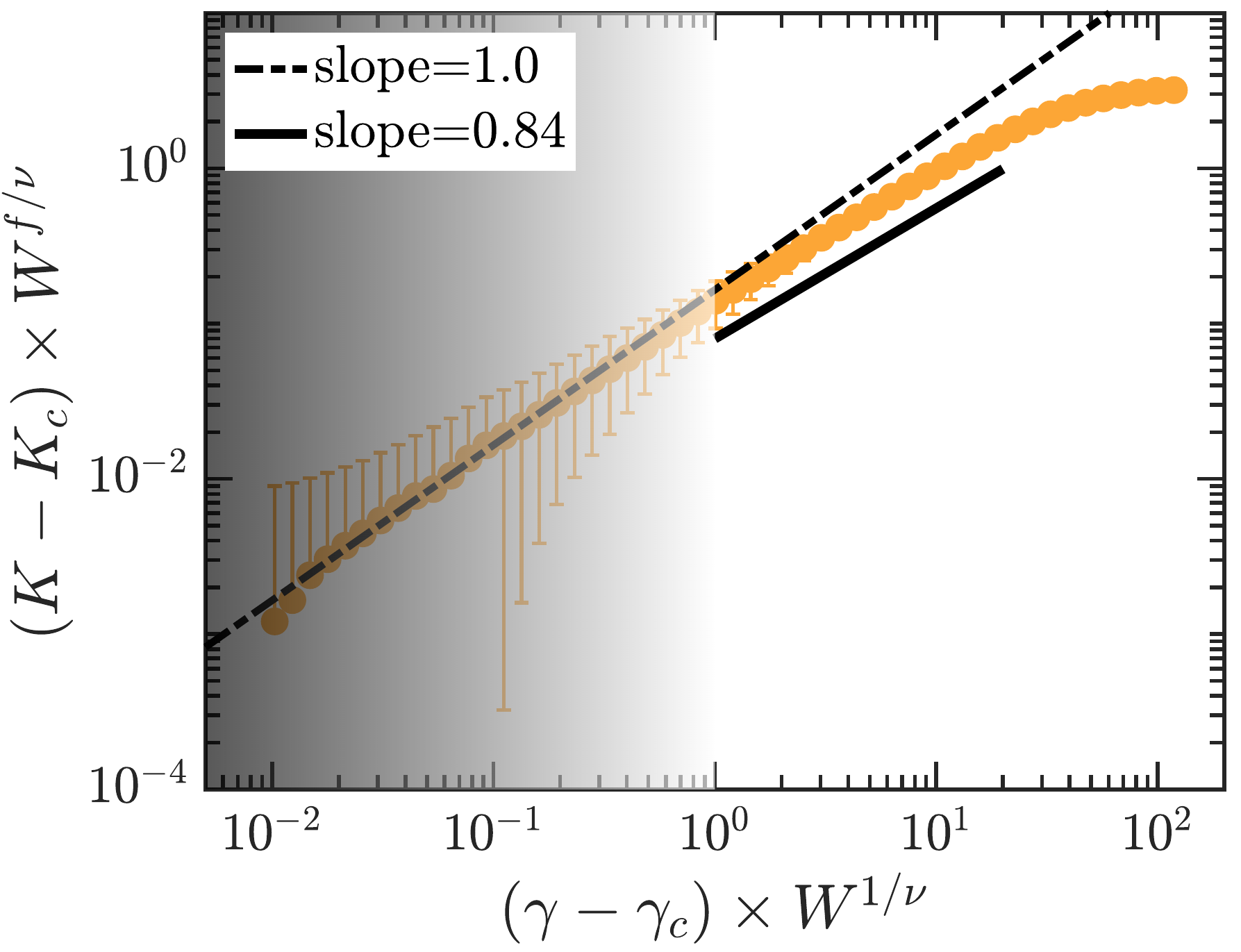}
	\caption{ \label{fig:A10} Finite-size effects for a 3D packing-derived network with $z=3.3$ and $W=20$. In the critical region, we find a non-mean-field exponent $f = 0.84$. The finite-size dominated region is shaded.}
\end{figure}
\FloatBarrier

\subsection*{The effect of $K_c$ on the exponent $f$}
\begin{figure}[!h]
	\includegraphics[width=14cm,height=14cm,keepaspectratio]{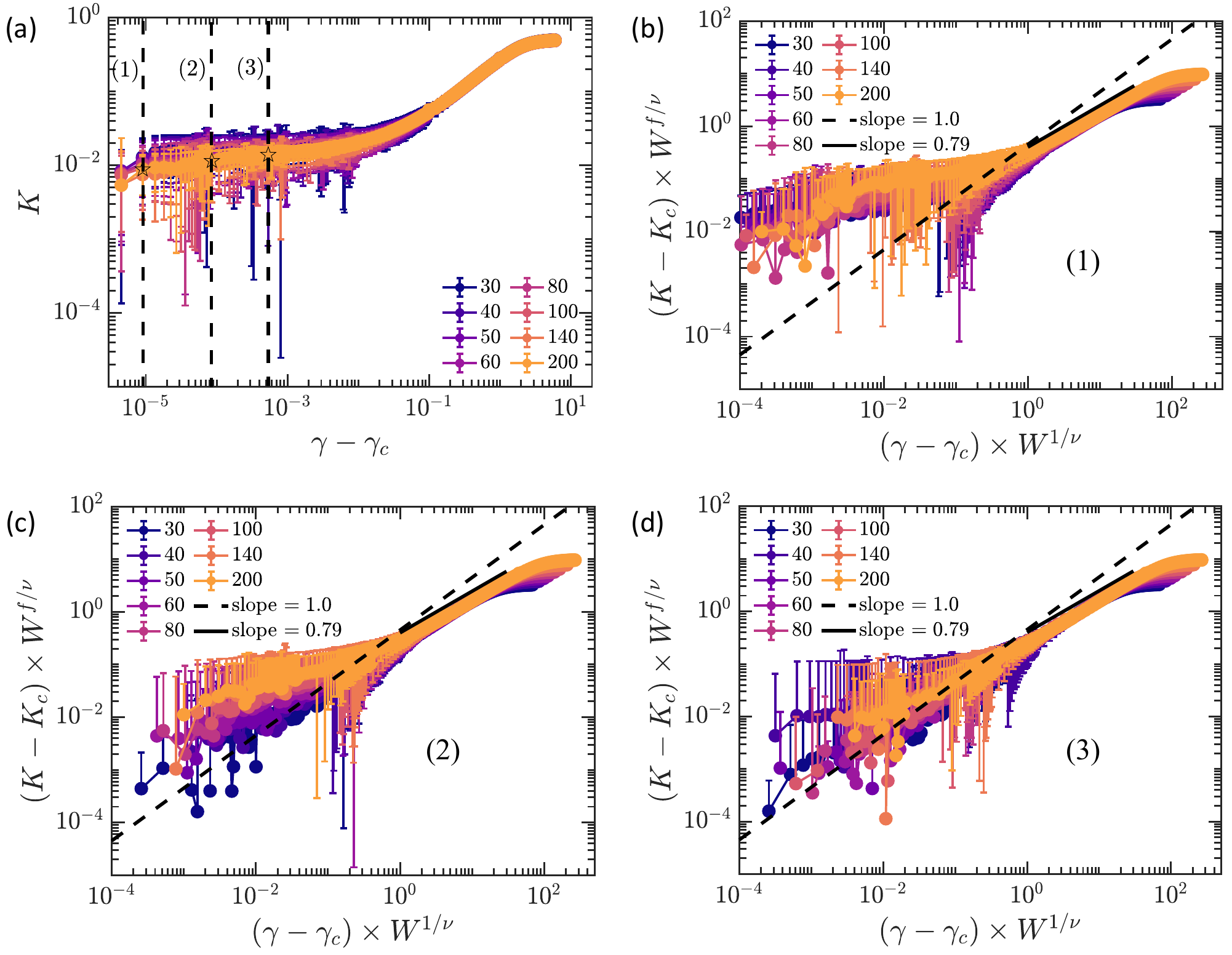}
	\caption{ \label{fig:A11} (a) Differential shear modulus versus $\gamma - \gamma_c$ for triangular networks with $z=3.3$. Plots (b)-(d) show the scaling analysis of the data in (a) using $K_c$ values corresponding to $\gamma - \gamma_c$ at vertical lines (1)-(3) in plot (a).}
\end{figure}

The scaling exponent $f$, which is obtained in the critical regime, is robust to errors in the value of discontinuity $K_c$. Figure\ \ref{fig:A11} shows that choosing different values for $K_c$ in a triangular network has negligible effect on $f$. Although the jammed-packing-derived model exhibits a slope of $1.0$ in the finite-size dominated region, the triangular model behaves differently (see Fig.\ 9). This is due to the fact that in contrast to packing-derived networks, triangular networks are likely to be rigidified by a single straight path of bonds connecting upper and lower boundaries of the simulation box in the small strain regime. Therefore, the $K_c$ values for a triangular network that are observed for small strains are results of these strand-like tensions. As we increase the strain, more bonds become involved, thus the slope in the finite-size dominated region gets closer to $1.0$, similar to packing-derived networks. This is clearly observed by choosing different $K_c$ values for finite-size scaling analysis of triangular networks (see Fig.\ \ref{fig:A11}).

\FloatBarrier

\begin{figure}[!h]
	\includegraphics[width=8.5cm,height=8.5cm,keepaspectratio]{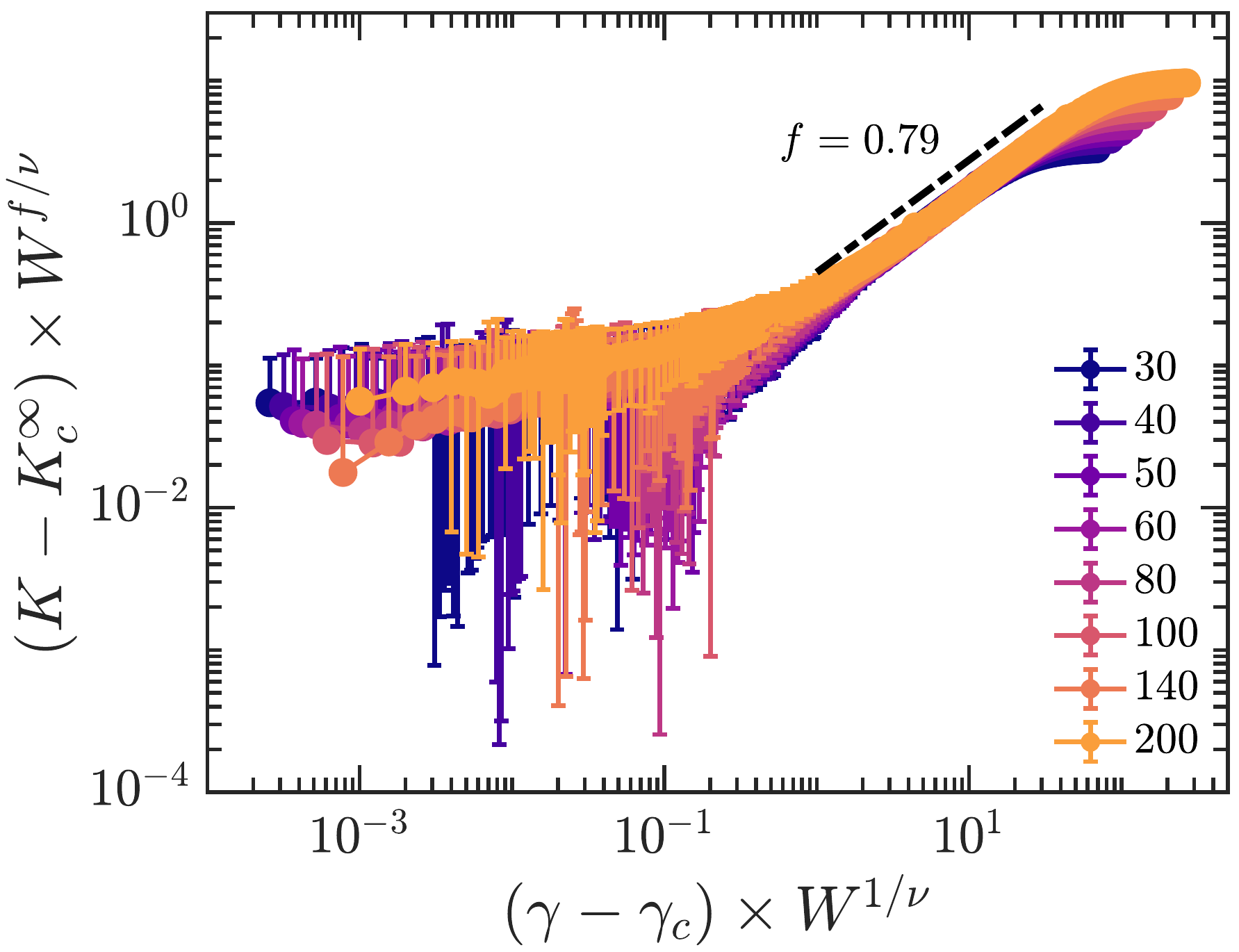}
	\caption{ \label{fig:A12} Finite-size scaling of the data in Fig.\ 9 a in the main text, using $K_c$ in the thermodynamic limit.}
\end{figure}

By using the modulus discontinuity in the thermodynamic limit $K_c^{\infty}$, we repeat the analysis performed in Fig.\ 9 a in the main text. As can be observed in Fig.\ \ref{fig:A12}, we find the same non-mean-field scaling exponent $f$.

\FloatBarrier

\subsection*{Fiber networks with bending interactions}

Using central force networks, we are only able to investigate the positive side of the transition, i.e., $\gamma-\gamma_c \rightarrow 0^{+}$. In order to understand the system's behavior below the critical point, we stabilize the networks by introducing weak bending interactions between bonds. Therefore, the elastic energy for the network has both stretching $E_s$ and bending $E_b$ contributions

\begin{equation}\label{}
E = E_s + E_b = \frac{\mu}{2} \sum_{ij}^{}\frac{(\ell_{ij} - \ell_{ij,0})^2}{\ell_{ij,0}} +
\frac{\kappa}{2} \sum_{ij}^{}\frac{(\theta_{ijk} - \theta_{ijk,0})^2}{\ell_{ijk,0}},
\end{equation}

in which the stretching part $E_s$ is the same as in Eq.\ 1 in the main text, $\kappa$ is the bending stiffness of individual fibers, $\theta_{ijk,0}$ is the angle between bonds $ij$ and $jk$ in the undeformed state, $\theta_{ijk}$ is the angle between those bonds after deformation, and $\ell_{ijk,0} = \frac{1}{2} (\ell_{ij,0} + \ell_{jk,0})$. Note that the bending energy is defined for consecutive bonds along each fiber on the triangular lattice. In simulations, we set $\mu=1$ and vary the dimensionless bending stiffness $\tilde{\kappa} = \kappa/\mu\ell_0^2$, where $\ell_0$ is the typical bond length ($\ell_0=1$ in lattice models).

\begin{figure}[!h]
	\includegraphics[width=18cm,height=18cm,keepaspectratio]{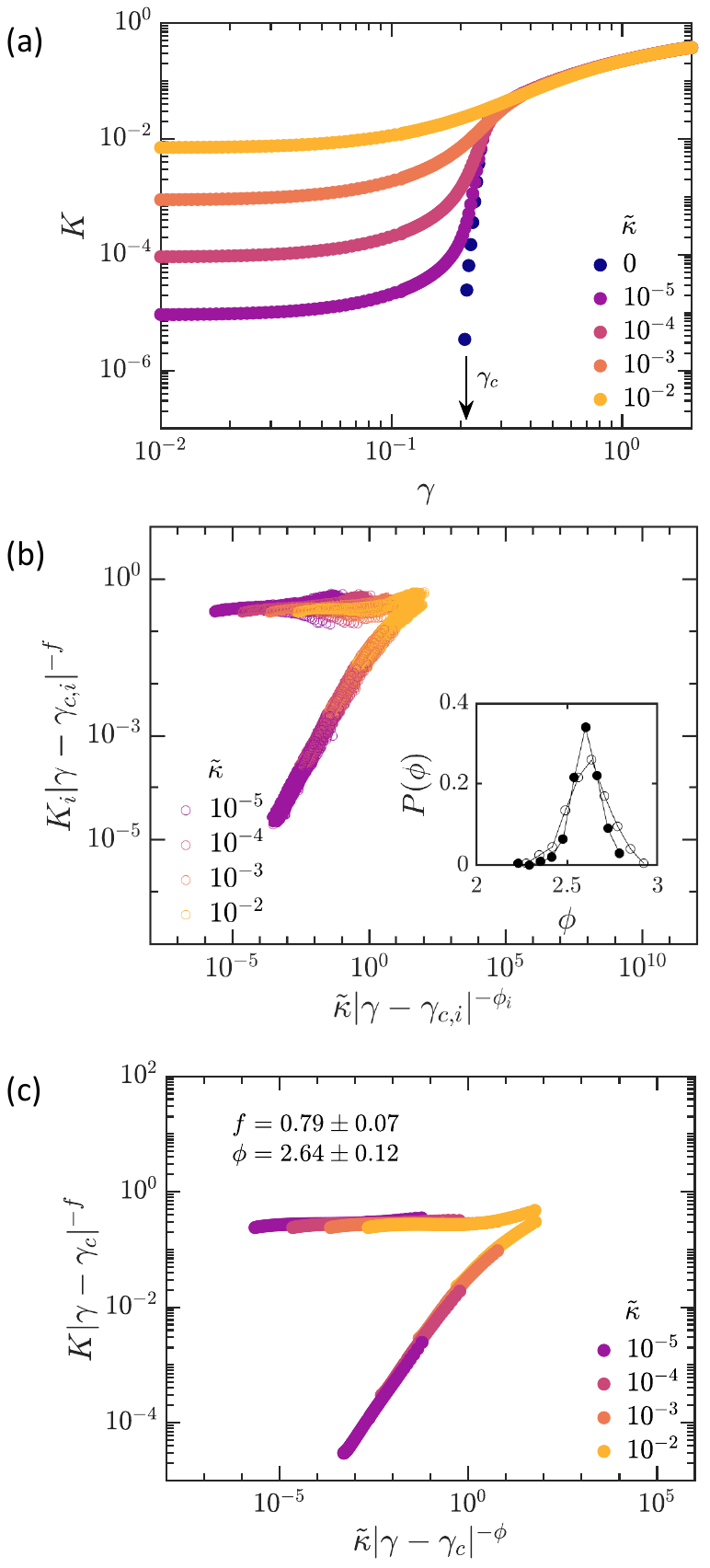}
	\caption{ \label{fig:A13} (a) The differential shear modulus versus strain for triangular networks with $W=100, z=3.3$  and varying the dimensionless bending rigidity $\tilde{\kappa}$. (b) The Widom-like collapse of individual samples in (a) according to Eq.\ 7 in the main text using the exponent $f$ that is already obtained for central force networks. Note that the finite-size-dominated data in which $|\Delta \gamma| \times W^{1/\nu} < 1.0$ are removed from this plot. Inset: showing the distribution of $\phi$, which are collected in $\gamma < \gamma_c$ regime of Eq.\ 7 in the main text. The $\phi$ values here are obtained using data with $\tilde{\kappa} = 10^{-5}$. The solid symbols are corresponding to $\phi$ values obtained using the ensemble average $f$, the empty symbols, on the other hand, are the distribution of $\phi$ exponents that collected using sample-specific $f$. (c) The Widom-like collapse similar to (b), but for the ensemble average of data. We note that the finite-size-dominated data in which $|\Delta \gamma| \times W^{1/\nu} < 1.0$ are removed from this plot.}
\end{figure}

The simulation procedure for networks with bending interactions is basically the same as discussed in the main text for central force networks. The differential shear modulus $K$ versus shear strain is shown in Fig.\ \ref{fig:A13} a for various dimensionless bending rigidity $\tilde{\kappa}$.
\FloatBarrier

\end{document}